\documentclass[notitlepage,twocolumn,letterpaper,natbib,aps,prl,amsmath,amsfonts,nofootinbib,preprintnumbers,superscriptaddress,secnumarabic]{revtex4-1}
\pdfoutput=1
\usepackage{amssymb,amsmath,latexsym,mathrsfs}
\usepackage{url}
\usepackage{enumitem}
\usepackage{graphicx}
\usepackage{booktabs}
\usepackage[usenames,dvipsnames]{color}
\usepackage[breaklinks,colorlinks,urlcolor=magenta,citecolor=magenta,linkcolor=magenta]{hyperref}
\usepackage{multirow}
\usepackage{float}
\usepackage{cases}
\usepackage{blindtext}
\usepackage{pifont}
\usepackage{hhline}
\usepackage{xcolor}
\definecolor{linkcolor}{rgb}{0.0, 0.47, 0.75}
\definecolor{citecolor}{rgb}{1.0, 0.5, 0.0}
\hypersetup{
  linkcolor  = linkcolor,
  citecolor  = linkcolor,
  urlcolor   = linkcolor,
  colorlinks = true
}

\makeatletter
\def\maketitle{
\@author@finish
\title@column\titleblock@produce
\suppressfloats[t]}
\makeatother

\begin{document}

\title{What to do when things get crowded? \\ Scalable joint analysis of overlapping gravitational wave signals}

\author{James Alvey${}^\ddagger$}
\email{j.b.g.alvey@uva.nl}
\thanks{ORCID: \href{https://orcid.org/0000-0003-2020-0803}{0000-0003-2020-0803}}
\affiliation{GRAPPA Institute, Institute for Theoretical Physics Amsterdam,\\
University of Amsterdam, Science Park 904, 1098 XH Amsterdam, The Netherlands}

\author{Uddipta Bhardwaj${}^\ddagger$}
\email{u.bhardwaj@uva.nl}
\thanks{ORCID: \href{https://orcid.org/0000-0003-1233-4174}{0000-0003-1233-4174}}
\affiliation{GRAPPA Institute, Anton Pannekoek Institute for Astronomy and Institute of High-Energy Physics,\\
University of Amsterdam, Science Park 904, 1098 XH Amsterdam, The Netherlands}

\author{\\Samaya Nissanke}
\affiliation{GRAPPA Institute, Anton Pannekoek Institute for Astronomy and Institute of High-Energy Physics,\\
University of Amsterdam, Science Park 904, 1098 XH Amsterdam, The Netherlands}
\affiliation{Nikhef, Science Park 105, 1098 XG Amsterdam, The Netherlands}

\author{Christoph Weniger}
\affiliation{GRAPPA Institute, Institute for Theoretical Physics Amsterdam,\\
University of Amsterdam, Science Park 904, 1098 XH Amsterdam, The Netherlands}

\begin{abstract}

\noindent The gravitational wave sky is starting to become very crowded, with the fourth science run (O4) at LIGO expected to detect $\mathcal{O}(100)$ compact object coalescence signals. Data analysis issues start to arise as we look further forwards, however. In particular, as the event rate increases in \textit{e.g.} next generation detectors, it will become increasingly likely that signals arrive in the detector coincidentally, eventually becoming the dominant source class. It is known that current analysis pipelines will struggle to deal with this scenario, predominantly due to the scaling of traditional methods such as Monte Carlo Markov Chains and nested sampling, where the time difference between analysing a single signal and multiple can be as significant as days to months. In this work, we argue that sequential simulation-based inference methods can solve this problem by breaking the scaling behaviour. Specifically, we apply an algorithm known as (truncated marginal) neural ratio estimation (TMNRE), implemented in the code \texttt{peregrine} and based on \texttt{swyft}. To demonstrate its applicability, we consider three case studies comprising two overlapping, spinning, and precessing binary black hole systems with merger times separated by 0.05 s, 0.2 s, and 0.5 s. We show for the first time that we can recover, with full precision (as quantified by a comparison to the analysis of each signal independently), the posterior distributions of all 30 model parameters in a full joint analysis. Crucially, we achieve this with only $\sim 15\%$ of the waveform evaluations that would be needed to analyse even a single signal with traditional methods.

\vspace*{5pt} \noindent \textbf{\texttt{GitHub}}: The \texttt{peregrine} analysis and inference library is available \href{https://github.com/PEREGRINE-GW/peregrine/tree/overlapping}{here} (\texttt{peregrine-gw/peregrine}). In addition, the TMNRE implementation \texttt{swyft} is available \href{https://github.com/undark-lab/swyft}{here} (\texttt{undark-lab/swyft}). \\ {\small${}^\ddagger$These two authors contributed equally to this work.}
\end{abstract}

\maketitle
\hypersetup{
  linkcolor  = linkcolor,
  citecolor  = linkcolor,
  urlcolor   = linkcolor
}

\noindent \emph{Introduction.---} The advent of gravitational wave (GW) detection~\cite{LIGOScientific:2016aoc} has already revolutionized our understanding of the Universe, providing a powerful tool for studying theories of gravity~\cite{LIGOScientific:2020tif}, astrophysics~\cite{LIGOScientific:2020kqk,LIGOScientific:2018cki}, and cosmology~\cite{LIGOScientific:2019zcs,Nissanke:2009kt,Nissanke:2011ax}. Up to now, over 90 compact binary coalescences (CBCs) have been detected globally, and live observations by the LIGO-Virgo-Kagra Collaboration (LVKC) in its fourth observing run currently add new observations on an almost daily basis~\cite{LIGOScientific:2020ibl,LIGOScientific:2021djp,LIGOScientific:2021usb,Nitz:2021uxj,Venumadhav:2019aa,Venumadhav:2020aa}. Looking forwards, future enhancements to the detector network include the introduction of KAGRA~\cite{KAGRA:2013rdx, KAGRA:2020tym}, LIGO-India~\cite{IndiGO:2011aaa, Unnikrishnan:2013qwa}, and the proposed LIGO-voyager~\cite{LIGO:2020xsf}, alongside the next-generation detectors such as the Einstein Telescope (ET)~\cite{Maggiore:2019uih}, Cosmic Explorer (CE)~\cite{Reitze:2019iox}, and the space-based Laser Interferometer Space Antenna (LISA)~\cite{LISA:2017aaa}. Combined, these promise to significantly expand our coverage of the gravitational wave landscape, both in terms of number and distinct class of event, as we probe new cosmic distances and frequency ranges.

In this context, however, the improved sensitivity and expanded coverage of these advanced detectors brings forth a number of concrete data analysis challenges. For example, these include the analysis of long signals from \textit{e.g.} large mass ratio inspirals, the impact of non-stationary or unknown detector noise characterisation on parameter inference, or the global analysis of a (set of) stochastic background and transient signals 
(see \textit{e.g.}~\cite{Littenberg:2023xpl, Babak:2017tow, Spadaro:2023muy, Weaving:2023fji, Sah:2023bgr, Pozzoli:2023kxy, Flauger:2020qyi, Bandopadhyay:2023gkb, Katz:2021yft}). In this work, we will focus specifically on an additional problem: the analysis of overlapping binary coalescence signals within a single detection window~\cite{Pizzati:2021apa, Antonelli:2021vwg,Samajdar:2021egv,Janquart:2022nyz,Langendorff:2022fzq}. Although this issue is not necessarily directly relevant to the current LVKC observing run (although it is possible given the detector upgrades that such an event could be observed in O4/5), the presence of large numbers ($\mathcal{O}(1000)s$) of overlapping signals in future detector networks is overwhelmingly likely~\cite{Relton:2021cax}. More precisely, current estimates of merger rates~\cite{LIGOScientific:2020kqk} have been used to predict the probability of detecting concurrent CBCs in multiple current and next generation configurations, see \textit{e.g.}~\cite{Relton:2021cax, Relton:2022whr}. These references have reported that observing overlapping CBCs in a 3G detector network is near certain. Another way to state the problem is the following: given current merger rate estimates, suppose an event is above the detectable threshold in a next generation detector network, the probability that there is an additional overlapping source in the same detection window is effectively~$100\%$.

The presence of this dominant class of overlapping signals presents a significant challenge for the applicability of current parameter inference pipelines (\textit{e.g.}~\cite{Ashton:2018jfp, Romero-Shaw:2020owr, Biwer:2018osg, Veitch:2014wba}) in terms of both computational feasibility, as well as statistical robustness. 
\begin{figure}[ht]
    \centering
    \includegraphics[width=\columnwidth,trim={0.1cm 0.1cm 0.1cm 0.1cm},clip]{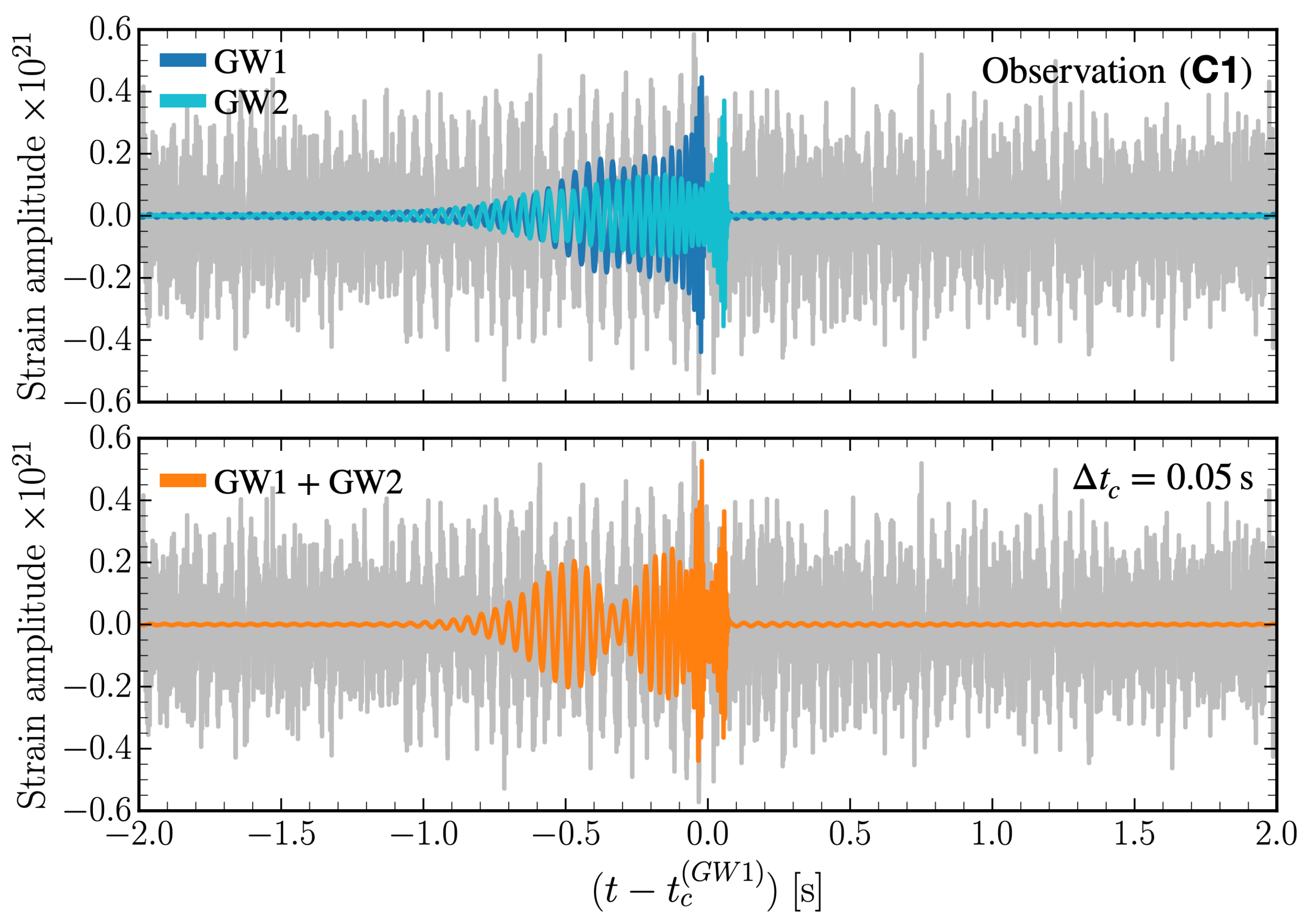}
    \caption{Time domain representation of two spinning, precessing BBH gravitational wave signals overlapping with $\Delta t_c = 0.05\,\mathrm{s}$. Depicting one of our case studies (\textbf{C1}), the top panel shows the individual component signals \textbf{GW1} (with $M_1=42.2\,M_\odot,\,M_2=32.5\,M_\odot$) and \textbf{GW2} (with $M_1=34.2\,M_\odot,\, M_2=27.7\,M_\odot$), both at a network signal-to-noise ratio of $20$. The signals are chronologically ordered according to merger times in LIGO-Hanford detector noise (grey). The bottom panel shows the overlapping signal (orange) resulting from the interference of \textbf{GW1} and \textbf{GW2}.} 
    \label{fig:C1_observation}\vspace{-10pt}
\end{figure}
There are a number of current strategies proposed in the literature as to how to approach this challenge. One possible method is to assume that exactly one of the signals is dominant, and then apply current parameter inference techniques such as Monte Carlo Markov Chains (MCMC)~\cite{Foreman-Mackey:2012any,Vousden:2016aaa} or nested sampling~\cite{Skilling:2006gxv,Handley:2015aaa,Ashton:2022grj,Speagle:2020aaa} with an individual waveform template. However, this has been shown to induce significant biases in the parameter reconstruction due to the effective mismodeling of the underlying noise realisation,\footnote{This is particularly pertinent for next generation detectors where there will be very few periods without signal in-band~\cite{Samajdar:2021egv}.} especially when the signals have similar merger times and/or frequency evolution.\footnote{Previous studies using Fisher forecasting have additionally highlighted that these biases are more pronounced for binary black hole (BBH) systems~\cite{Pizzati:2021apa}, owing to there being fewer in-band cycles compared to binary neutron star (BNS) and neutron star-black hole (NSBH) systems.} This motivated additional efforts to develop joint parameter estimation techniques through \textit{e.g.} modifications to the single waveform likelihood function to account for multiple overlapping CBCs (see \textit{e.g.} Ref.~\cite{Janquart:2022nyz}), or working directly with the joint likelihood. 

At a fundamental level, however, since traditional sampling algorithms such as MCMC and nested sampling generate samples directly from the full joint posterior, these algorithms tend to scale poorly as the dimensionality of the problem increases. For example, whilst an analysis of a single signal LVKC event may take $\mathcal{O}(1)\,\mathrm{day}$ with state of the art waveform templates, it has been shown that analysing even two waveforms could take $\mathcal{O}(1)$ month~\cite{Janquart:2022nyz}\footnote{In Ref.~\cite{Janquart:2022nyz}, the reported time taken for a joint parameter estimation using \texttt{dynesty}~\cite{Speagle:2020aaa} with a $30\,\mathrm{D}$ likelihood function is $23.8$ days on $16$ \textit{Intel(R) Xeon(R) Gold 6152} CPUs.}. This renders the implementation of such techniques for the potential $\mathcal{O}(10^5)$ events~\cite{Relton:2021cax} in a single observing run with 3G detectors computationally infeasible. It may be possible to improve this outlook with efficient high performance computing techniques or via the implementation of rapid waveform modelling techniques~\cite{Edwards:2023sak}, fast stochastic samplers~\cite{Wong:2023lgb} and relative binning~\cite{Zackay:2018qdy, Leslie:2021ssu}.

Regardless, these challenges motivate the exploration of alternative data analysis strategies. In this work, we will argue for the benefits of modern simulation-based inference (SBI) techniques (see \textit{e.g.} Ref.~\cite{Cranmer:2019eaq} for a review) in analysing overlapping gravitational wave signals. In general, SBI methods solve the Bayesian inference problem of performing parameter inference on a set of model parameters $\boldsymbol{\theta}$ given some observation $\mathbf{x}_0$ by forward modelling the data $\mathbf{x}$ as a function of $\boldsymbol{\theta}$. For many applications (see \textit{e.g.} Refs.~\cite{Miller:2021hys,Tejero-cantero:2020aaa,Alsing:2019xrx,Cole:2021gwr,Montel:2022fhv,AnauMontel:2022ppb,Dimitriou:2022cvc,Karchev:2022xyn,Alvey:2023pkx}), this crucially does not require the evaluation (or specification) of an explicit likelihood, although there are other benefits that are also relevant to GW studies such as simulation efficiency, amortized inference, and the ability to do marginal inference on subsets of a model. Particularly for GW parameter inference, an implementation of SBI called neural posterior estimation (NPE) has been shown to impressively perform fully amortized inference on compact binary mergers~\cite{Dax:2021myb, Dax:2021tsq, Dax:2022pxd}. In the context of overlapping gravitational waves analysis, a normalizing flow-based approach has shown promising results~\cite{Langendorff:2022fzq} by obtaining wide, but (crucially) more accurate posteriors as compared to traditional likelihood-based approaches. Currently, this represents the state-of-the-art for parameter inference of overlapping gravitational wave signals.

\begin{figure*}[th]
    \centering
    \includegraphics[width=\textwidth,trim={0.4cm 0.4cm 0.4cm 0.4cm},clip]{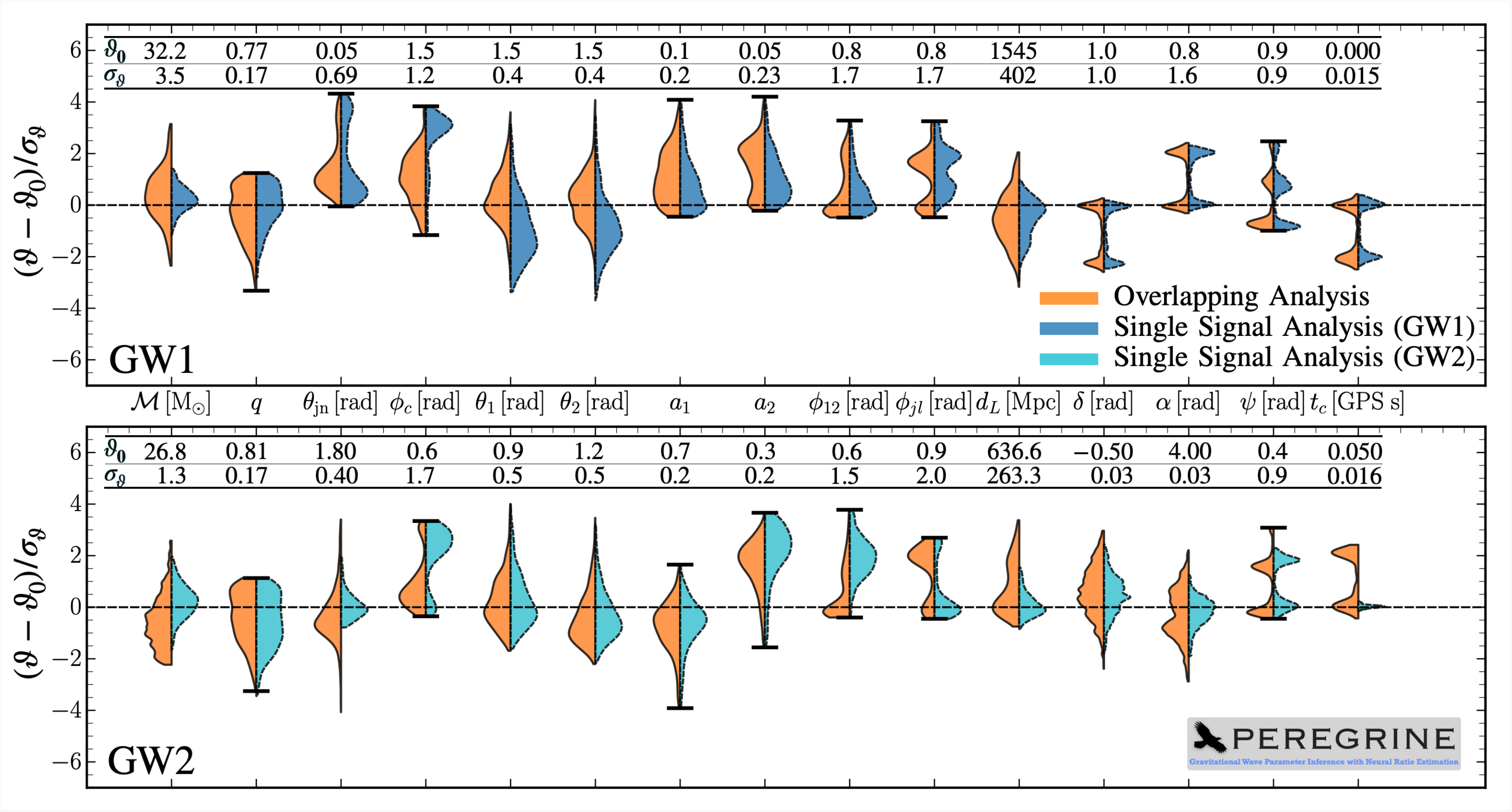}
    \caption{Violin plots showing $\mathrm{1D}$ marginal posterior distributions of all $30$ parameters characterising an overlapping signal comprising two concurrent BBH signals with $\Delta t_c = 0.05\,\mathrm{s}$ (as shown in Fig.~\ref{fig:C1_observation}). The left of each violin shows the $\mathrm{1D}$ marginal posterior obtained from our overlapping signal analysis (\textbf{C1}). The corresponding distributions on the right of each violin represent the posterior distributions obtained via a single signal analysis of each component signal in the absence of the other waveform. The top and bottom panels indicate parameters of the first (\textbf{GW1}) and second (\textbf{GW2}) signal respectively. To aid comparison across parameters, all $\mathrm{1D}$ posterior samples are scaled such that the injected value, $\vartheta_0$ is at $0$ (indicated by the dashed horizontal line), and normalized by the standard deviations of the overlapping posterior samples, $\sigma_\vartheta$. When shown, black horizontal lines at the ends of the posteriors indicate the (physical) edges of the prior boundary, \textit{e.g.} for the mass ratio $q$, the limits are at $q = 0, 1$. When not shown, this means that the posterior lies well within the initial prior range.
    }
    \label{fig:C1_violin_plots}\vspace{-10pt}
\end{figure*}

In the current work, we will make use of another SBI algorithm, known as (truncated marginal) neural ratio estimation, TMNRE~\cite{Miller:2021hys} to analyse overlapping GW signals. NRE functions by estimating the posterior-to-prior ratio via a binary classification task, see Refs.~\cite{Miller:2021hys,Miller:2022shs} for more information. Recently, this was implemented for the first time for the analysis of individual compact binary mergers in the code \texttt{peregrine}~\cite{Bhardwaj:2023xph}, built on top of \texttt{swyft}~\cite{Miller:2021hys,Miller:2022shs}. There are a number of key results from Ref.~\cite{Bhardwaj:2023xph} that suggest that this approach could solve the challenges facing overlapping analysis. Specifically, it was shown that \texttt{peregrine} could be used to analyse individual signals with only 2\% of the waveform evaluations required in a traditional approach such as nested sampling. The fundamental reason for why our approach is more simulation efficient is the fact that TMNRE focuses sequentially on lower dimensional marginal posteriors. In more detail, directly estimating (typically $\mathrm{1D}$ and $\mathrm{2D}$) marginal posteriors has been shown to be extremely efficient in a number of contexts, see e.g.~\cite{Cole:2021gwr,AnauMontel:2022ppb}. The sequential aspect of the algorithm is achieved through a prior truncation scheme that composes well with marginalization~\cite{Miller:2021hys} carried out over a series of inference rounds, and allows us to quickly zoom in to the relevant region of parameter space for a given observation. For example, the geocentric merger time below (see bottom panel of Fig.~\ref{fig:C1_violin_plots}) is constrained to within $\sim$1\% of the initial prior range, so truncating to this region can significantly reduce the data variance. For more details on both of these aspects, see Refs.~\cite{Miller:2021hys,Miller:2022shs,Bhardwaj:2023xph}. In the present context, assuming that the sort of simulation efficiency illustrated in Ref.~\cite{Bhardwaj:2023xph} carries through to the overlapping case -- which we will demonstrate below -- the inference time could be brought down from $\mathcal{O}$(1 month)~\cite{Janquart:2022nyz} to $\mathcal{O}$(1 day).

In the rest of the work, we will begin by setting up the case study that we will analyse -- two overlapping spinning, precessing binary black hole signals that have similar signal-to-noise ratios in a detector with LIGO and Virgo O4 (current) design sensitivity. Then, we will demonstrate that we can achieve full precision and accuracy in the parameter inference across the full 30-dimensional parameter space using the inference results for the individual signals as an optimality test. Finally, we provide some discussions and conclusions regarding our approach and results. Additional technical details regarding the \texttt{peregrine} implementation can be found in either the \textit{Supplemental Material} below, or in Ref.~\cite{Bhardwaj:2023xph}.  

\vspace{8pt}
\noindent \emph{An Overlapping Case Study.---} To explore the challenges posed by overlapping signals, we have designed three case studies with varying degrees of overlap as specified by the merger times, set at $\Delta t_c = 0.05 \, \mathrm{s}$ (\textbf{C1}), $0.2 \, \mathrm{s}$ (\textbf{C2}) and $0.5 \, \mathrm{s}$ (\textbf{C3}). All of these choices fall within the high-bias regime, as defined in \textit{e.g.}~\cite{Pizzati:2021apa}.
In each case study, we select two spinning, precessing binary black hole (BBH) signals with similar mass ratios and chirp masses. This choice ensures that the frequency evolution of the sources is comparable, allowing for extended periods of overlap in both time and frequency domains. Moreover, we order the mergers such that the lighter mass binary merges after the heavier mass binary, aiming at maximum overlap of their corresponding signals.
Furthermore, in order to eliminate the dominance of a single signal, we set the network signal-to-noise ratio (SNR) of the individual component signals to be the same at a value of $20$. This is achieved by appropriately scaling the luminosity distances of the signals once we have fixed the underlying noise realization. By ensuring similar SNRs in addition to the aforementioned setup, we simulate the most challenging scenario for parameter estimation of this specific class of signals. We provide an example observation in Fig.~\ref{fig:C1_observation}, which corresponds to one of our case studies (\textbf{C1}). The figure shows overlapping GW signals in the time domain with a merger time separation of $\Delta t_c = 0.05$ s. More specifically, the signals are generated using source parameters detailed in Tab.~\ref{tab:sbi_params} (\textit{Supplemental Material}) and injected onto a three-detector network comprising of the two LIGO detectors and Virgo.\footnote{We note that for next generation detectors, the substantial increase in bandwidth will lead to a significantly higher number of cycles in-band for such signals~\cite{Maggiore:2019uih,Reitze:2019iox}, which may improve the outlook.}

As validation of our results, we perform a separate analysis of the individual component signals using \texttt{peregrine} and the approach detailed in Ref.~\cite{Bhardwaj:2023xph}, where it was shown that the TMNRE approach~\cite{Miller:2021hys,Miller:2022shs} is highly efficient in achieving excellent agreement with e.g. MCMC. This allows us to refer to the parameter estimation performance for non-overlapping signals and establish a baseline for comparison. This is an important step forward in terms of the state-of-the-art, where we look to move towards full precision as opposed to simply unbiased coverage, as was impressively demonstrated in Ref.~\cite{Janquart:2022nyz}. We emphasize that the comparison to the full overlapping analysis should not be interpreted in terms of ``superimposed" posteriors but rather as a means to understand the loss of precision introduced by signal overlap and the corresponding reduction in SNR.

For data generation, we consider a network comprising three detectors: the two LIGO detectors (Hanford and Livingston) and Virgo\footnote{In the current analysis, we do not consider the KAGRA detector, since the sources are at the edge of its projected sensitivity range~\cite{KAGRA:2013rdx}.}, all operating at the sensitivity level of the O4 observing run with a minimum frequency of $20$ Hz and a sampling rate of $2048$ Hz. To simulate detector noise, we use the LIGO and Virgo O4 power spectral density (PSD) curve~\cite{Ashton:2021anp,LIGOScientific:2019hgc}.
The GW signals in our simulations are evaluated using the state-of-the-art \texttt{IMRPhenomXPHM}~\cite{Pratten:2020ceb} waveform approximant. This waveform model is chosen due to its ability to accurately account for higher-order modes and precession effects\footnote{Studies involving harmonic decomposition of precessing waveforms have demonstrated how the effects of precession on the gravitational wave signal can mimic the beating of two waveforms overlapping and lead to degeneracies in parameter inference~\cite{Green:2021stw,Kim:2023scq}.}. It is worth noting that alternative waveform approximants, such as ones from the \texttt{SEOBNR} (\textit{e.g.}~\cite{Taracchini:2013rva, Bohe:2016gbl, Ossokine:2020kjp}) and \texttt{IMRPhenom} (\textit{e.g.}~\cite{Khan:2015jqa, Pratten:2020ceb, Hannam:2013oca}) families, can also be readily employed in our setup depending on the specific requirements and objectives of the analysis.

\begin{figure}
    \centering
    \includegraphics[width=\columnwidth,trim={0.1cm 0.1cm 0.1cm 0.1cm},clip]{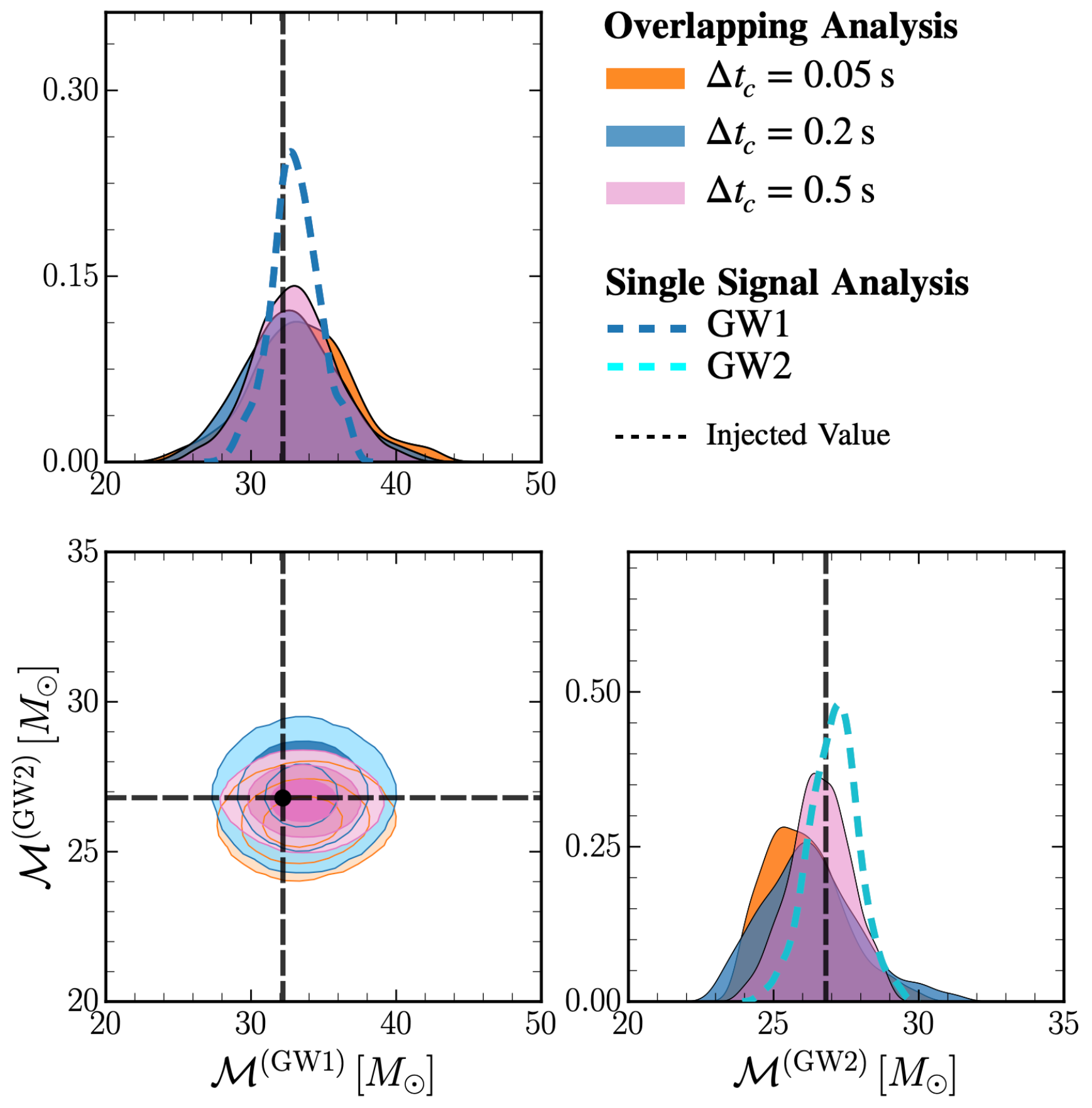}
    \caption{Corner plot showing the $\mathrm{1D}$ marginal posteriors for the chirp masses of the two BBH signals overlapping with three different degrees of time-overlap ($\Delta t_c = 0.05\,\mathrm{s}$, $0.2\,\mathrm{s}$, $0.5\,\mathrm{s}$ as shown in the figure legend).
    The $\mathrm{2D}$ marginal posteriors are shown with $\mathrm{1\sigma}, \mathrm{2\sigma}$ and   $\mathrm{3\sigma}$ contours indicated by subsequently lighter shading. The dashed $\mathrm{1D}$ marginal posteriors indicate the maximum precision and accuracy with which one can expect to constrain the parameters, as determined by analysing the individual signals.}
    \label{fig:corner_chirp_masses}
\end{figure}

\vspace{8pt}
\noindent \emph{Key Results with} \texttt{peregrine}\emph{.---} All of the key results in this work are in the context of the case studies described above. To summarise them briefly, our main discussion points are highlighted in Fig.~\ref{fig:C1_violin_plots} and Fig.~\ref{fig:corner_chirp_masses}. They illustrate two general points: our ability to achieve high precision and accuracy in our reconstruction of the posteriors for all 30 model parameters (15 for each signal), as well as our ability to investigate correlations between the two signals and the implications regarding the separation of the two waveforms in terms of precision and coverage. We discuss these points in more detail below, as well as the additional validation tests that we have carried out to support our results, and the computational efficiency of our algorithm in the present context. Full sets of detailed posteriors and coverage tests can be found in the \emph{Supplemental Material} (see Figs.~\ref{fig:2D_full}--~\ref{fig:C3_pp_full}).

The first result to discuss in detail is summarised in Fig.~\ref{fig:C1_violin_plots}. This highlights the comparison\footnote{The figure shows each $\mathrm{1D}$ marginal posterior rescaled with respect to its injection value $\vartheta_0$ and normalized by the standard deviation of the overlapping posterior samples $\sigma_\vartheta$ to ease comparison across parameters.} between the $\mathrm{1D}$ posteriors for all model parameters from our joint analysis of the overlapping signal in case study (\textbf{C1}, $\Delta t_c = 0.05\,\mathrm{s}$) as compared to the result for the corresponding individual signal. There are a number of things to comment on: firstly, we see that we achieve excellent coverage of the true injection value, represented by the horizontal dashed line. Secondly, and arguably the most important result of this work, we achieve remarkably similar precision compared to the corresponding analysis of the individual signal.\footnote{The single signal analysis here really acts as a sensitivity baseline, \textit{i.e.} it should not be possible to exceed the constraining power of the analysis in the absence of the second signal.} The posteriors are slightly broader in the overlapping analysis (see \textit{e.g.} the detector-frame chirp mass $\mathcal{M}$ for an illustrative example), which is to be expected since the presence of an additional signal must act to slightly reduce the overall SNR of the waveform under consideration. Nonetheless, we show for the first time that it is possible to obtain accurate parameter constraints with little to no loss of information. This should be compared to \textit{e.g.} the results in Ref.~\cite{Janquart:2022nyz} which achieved impressive unbiased coverage, but at the cost of significantly broader posteriors. For completeness, we comment on the one posterior that shows a distinct difference in sensitivity compared to our individual signal analysis baseline. Specifically, if one looks at the posteriors for the merger time of the second signal in the bottom panel of Fig.~\ref{fig:C1_violin_plots}, we see that the overlapping analysis results in a bimodal, and broader posterior on $t_c$. We believe this is fully explained by looking at the corresponding merger time of the first signal (\textbf{GW1}). The presence of two distinct modes (also evident in the individual-signal analysis) leads to a direct degeneracy between the merger time of the first and second signal that is fundamentally not present when analysing only the latter (\textbf{GW2}). Regardless, it is still extremely well measured as compared to the initial prior range (see Tab.~\ref{tab:sbi_params}).

We now focus attention on Fig.~\ref{fig:corner_chirp_masses} which presents the $\mathrm{1D}$ and $\mathrm{2D}$ posteriors for the chirp masses of each source (\textbf{GW1} and \textbf{GW2}) for all three case studies. We see again the level of precision that we achieve compared to the individual signal analysis, now compared across the different merger time differences. The general trend is clear that there is a slight, but noticeable, improvement in sensitivity on the chirp masses as the signals are separated. This is more prevalent in the case of \textbf{GW2}, where indeed for $\Delta t_c = 0.5\,\mathrm{s}$ it actually matches that of the individual signal. This brings up an interesting discussion regarding the impact of the underlying noise realisation on the precise position and shape of the posterior. Specifically, one might wonder why we do not exactly recover the posterior of the individual signal. We believe that the slight relative shift is a result of the (sub-dominant) presence of the ringdown of the first signal close to the merger of the second signal which effectively modifies slightly the underlying noise realisation. The result of this should exactly be a small observed shift in the posterior. The second distinct point to make about these results is the behaviour of the $\mathrm{2D}$ contours. We see that, as expected, they shrink as the merger times move further apart, but also that the two chirp masses are broadly uncorrelated. This is expected physically since the signals are totally distinct, however, it is also highlighting the fact that this measurement is not noise-dominated which could lead to induced degeneracies between the parameters. We believe this is the first time this has been explicitly shown for this type of overlapping analysis, and paints a hopeful picture for our ability to obtain high quality reconstructions of coincident signals in future detectors.

Although the individual signal analysis is a strong indication that our method is performing well, it is important to carry out additional validation tests. Coverage tests, see \textit{e.g.}~\cite{Hermans:2021aaa,Lemos:2023aaa} are now standard practice in simulation-based inference applications. The full outcome of these tests are shown in the \textit{Supplemental Material} in Figs.~\ref{fig:C1_pp_full}--\ref{fig:C3_pp_full}. The takeaway message from our analysis is that for all parameters, across all case studies, our posteriors are extremely well calibrated (lie along the diagonal as plotted in the (empirical coverage)-(expected coverage) plane).

The final important discussion regarding our analysis concerns computational efficiency. It has been explicitly shown that using traditional methods to carry out this type of joint inference is extremely costly, taking on average over 3 weeks to run (see Footnote 3 in Ref.~\cite{Janquart:2022nyz}). Taking our analysis of the first case study (\textbf{C1}) as a concrete example, we will make a direct comparison with this state-of-the-art joint inference benchmark. Across 7 sequential rounds of inference using \texttt{peregrine} (for details regarding our TMNRE implementation see Ref.~\cite{Bhardwaj:2023xph}), we required only $7 \times 10^6$ overlapping waveform simulations to achieve our analysis results. This is only a factor of 10 more than we required to perform analysis on a single signal in Ref.~\cite{Bhardwaj:2023xph} for twice as many parameters. This therefore convincingly breaks the expected scaling of traditional methods solving this joint inference problem, and is an order of magnitude fewer waveform evaluations than is typically required to analyse even a single signal with traditional methods such as MCMC. In terms of time, we used an 18-core CPU cluster node to generate the $7 \times 10^6$ waveform evaluations, which took on average around 2-3 hours per round of 1 million examples. Training the network each round is currently the main time cost of the analysis, taking between 8 and 12 hours per round for $10^6$ simulations. Overall, this means our analysis takes around 48 hours to complete, depending on the specific hardware choice.\footnote{Specifically, we ran our waveform simulations on 18 (shared) \textit{Intel(R) Xeon(R) Platinum 8360Y} CPUs and network training on a single, shared \texttt{NVIDIA} A100 graphics card.} For more details on the network specification and training procedure, please see the section in the \textit{Supplemental Material}.

\vspace{8pt}
\noindent \emph{Conclusions and Outlook.---} In this work, we have presented the application of the sequential simulation-based inference method in Ref.~\cite{Bhardwaj:2023xph} to the analysis of overlapping gravitational wave signals. We showed for the first time that we can obtain full precision and accuracy over all spinning, precessing binary black hole parameters in a joint inference analysis. Our main results are presented in Figs.~\ref{fig:C1_violin_plots} and~\ref{fig:corner_chirp_masses} for the three case studies described above. Crucially, to do so, we required only $7 \times 10^6$ waveform evaluations, which is an order of magnitude fewer than traditional methods typically require to analyse \emph{even a single signal}.

As far as outlook is concerned, there are a number of important points to make regarding our method. The first is scalability: we have highlighted that we can perform full inference on two overlapping signals with a computational cost and simulation efficiency that is at least an order of magnitude lower than the current state-of-the-art joint inference methods (see \textit{e.g.} Refs.~\cite{Pizzati:2021apa,Antonelli:2021vwg,Samajdar:2021egv,Janquart:2022nyz,Langendorff:2022fzq}). There is significant room for improvement in this direction, for example, we could use pre-trained networks from earlier inference rounds and fine-tune them to significantly accelerate our training, utilise accelerated waveform models~\cite{Edwards:2023sak}, or implement a more efficient truncation scheme that accounts for parameter correlations. Secondly, if we look towards the science case of future experiments such as the Einstein Telescope~\cite{Maggiore:2019uih}, Cosmic Explorer~\cite{Reitze:2019iox}, and LISA~\cite{LISA:2017aaa}, overlapping signals (of various types) will be a dominant source class. In order to use our joint analysis framework for an entirely realistic next generation gravitational wave scenario, natural development is required in the sense of using 3G interferometer setups, considering multiple (\textit{i.e.} more than $2$) signal overlaps, carrying out the initial triggering of these overlapping events, and dealing with very long signals (for \textit{e.g.} next generation detection of binary neutron star signals). Another class of analysis problems in the future will be the characterisation of detector noise, since almost all of the data is expected to contain some signal component. This is in addition to the fact that the noise may potentially be auto-correlated. It will be interesting to investigate the simulation-efficiency achieved in such a next generation detector (ET/CE) setting. Partly as a result of the implicit likelihood nature of \texttt{peregrine}~\cite{Bhardwaj:2023xph} as a method, combined with its scalability, we believe it offers a flexible starting point for this development towards a data analysis pipeline suitable for the next generation of gravitational wave detectors.

\vspace*{-10pt}
\section*{Acknowledgements}
\noindent This work is part of the project CORTEX (NWA.1160.18.316) of the research programme NWA-ORC which is (partly) financed by the Dutch Research Council (NWO). Additionally, CW acknowledges funding from the European Research Council (ERC) under the European Union's Horizon 2020 research and innovation programme (Grant agreement No. 864035). UB is supported through the CORTEX project of the NWA-ORC with project number NWA.1160.18.316 which is partly financed by the Dutch Research Council (NWO). JA is supported through the research program ``The Hidden Universe of Weakly Interacting Particles" with project number 680.92.18.03 (NWO Vrije Programma), which is partly financed by the Nederlandse Organisatie voor Wetenschappelijk Onderzoek (Dutch Research Council). SN acknowledges support from the NWO Projectruimte grant (Samaya Nissanke). Additionally, CW acknowledges funding from the European Research Council (ERC) under the European Union's Horizon 2020 research and innovation programme (Grant agreement No. 864035). We thank Thomas Edwards and Pablo Bosch for useful conversations and comments on a draft of this manuscript. The main analysis for this work was carried out on the Snellius Computing Cluster at SURFsara.

\bibliography{biblio}

\begin{thebibliography}{78}%
\makeatletter
\providecommand \@ifxundefined [1]{%
 \@ifx{#1\undefined}
}%
\providecommand \@ifnum [1]{%
 \ifnum #1\expandafter \@firstoftwo
 \else \expandafter \@secondoftwo
 \fi
}%
\providecommand \@ifx [1]{%
 \ifx #1\expandafter \@firstoftwo
 \else \expandafter \@secondoftwo
 \fi
}%
\providecommand \natexlab [1]{#1}%
\providecommand \enquote  [1]{``#1''}%
\providecommand \bibnamefont  [1]{#1}%
\providecommand \bibfnamefont [1]{#1}%
\providecommand \citenamefont [1]{#1}%
\providecommand \href@noop [0]{\@secondoftwo}%
\providecommand \href [0]{\begingroup \@sanitize@url \@href}%
\providecommand \@href[1]{\@@startlink{#1}\@@href}%
\providecommand \@@href[1]{\endgroup#1\@@endlink}%
\providecommand \@sanitize@url [0]{\catcode `\\12\catcode `\$12\catcode
  `\&12\catcode `\#12\catcode `\^12\catcode `\_12\catcode `\%12\relax}%
\providecommand \@@startlink[1]{}%
\providecommand \@@endlink[0]{}%
\providecommand \url  [0]{\begingroup\@sanitize@url \@url }%
\providecommand \@url [1]{\endgroup\@href {#1}{\urlprefix }}%
\providecommand \urlprefix  [0]{URL }%
\providecommand \Eprint [0]{\href }%
\providecommand \doibase [0]{http://dx.doi.org/}%
\providecommand \selectlanguage [0]{\@gobble}%
\providecommand \bibinfo  [0]{\@secondoftwo}%
\providecommand \bibfield  [0]{\@secondoftwo}%
\providecommand \translation [1]{[#1]}%
\providecommand \BibitemOpen [0]{}%
\providecommand \bibitemStop [0]{}%
\providecommand \bibitemNoStop [0]{.\EOS\space}%
\providecommand \EOS [0]{\spacefactor3000\relax}%
\providecommand \BibitemShut  [1]{\csname bibitem#1\endcsname}%
\let\auto@bib@innerbib\@empty
\bibitem [{\citenamefont {Abbott}\ \emph {et~al.}(2016)\citenamefont {Abbott}
  \emph {et~al.}}]{LIGOScientific:2016aoc}%
  \BibitemOpen
  \bibfield  {author} {\bibinfo {author} {\bibfnamefont {B.~P.}\ \bibnamefont
  {Abbott}} \emph {et~al.} (\bibinfo {collaboration} {LIGO Scientific,
  Virgo}),\ }\href {\doibase 10.1103/PhysRevLett.116.061102} {\bibfield
  {journal} {\bibinfo  {journal} {Phys. Rev. Lett.}\ }\textbf {\bibinfo
  {volume} {116}},\ \bibinfo {pages} {061102} (\bibinfo {year} {2016})},\
  \Eprint {http://arxiv.org/abs/1602.03837} {arXiv:1602.03837 [gr-qc]}
  \BibitemShut {NoStop}%
\bibitem [{\citenamefont {Abbott}\ \emph
  {et~al.}(2021{\natexlab{a}})\citenamefont {Abbott} \emph
  {et~al.}}]{LIGOScientific:2020tif}%
  \BibitemOpen
  \bibfield  {author} {\bibinfo {author} {\bibfnamefont {R.}~\bibnamefont
  {Abbott}} \emph {et~al.} (\bibinfo {collaboration} {LIGO Scientific,
  Virgo}),\ }\href {\doibase 10.1103/PhysRevD.103.122002} {\bibfield  {journal}
  {\bibinfo  {journal} {Phys. Rev. D}\ }\textbf {\bibinfo {volume} {103}},\
  \bibinfo {pages} {122002} (\bibinfo {year} {2021}{\natexlab{a}})},\ \Eprint
  {http://arxiv.org/abs/2010.14529} {arXiv:2010.14529 [gr-qc]} \BibitemShut
  {NoStop}%
\bibitem [{\citenamefont {Abbott}\ \emph
  {et~al.}(2021{\natexlab{b}})\citenamefont {Abbott} \emph
  {et~al.}}]{LIGOScientific:2020kqk}%
  \BibitemOpen
  \bibfield  {author} {\bibinfo {author} {\bibfnamefont {R.}~\bibnamefont
  {Abbott}} \emph {et~al.} (\bibinfo {collaboration} {LIGO Scientific,
  Virgo}),\ }\href {\doibase 10.3847/2041-8213/abe949} {\bibfield  {journal}
  {\bibinfo  {journal} {Astrophys. J. Lett.}\ }\textbf {\bibinfo {volume}
  {913}},\ \bibinfo {pages} {L7} (\bibinfo {year} {2021}{\natexlab{b}})},\
  \Eprint {http://arxiv.org/abs/2010.14533} {arXiv:2010.14533 [astro-ph.HE]}
  \BibitemShut {NoStop}%
\bibitem [{\citenamefont {Abbott}\ \emph
  {et~al.}(2018{\natexlab{a}})\citenamefont {Abbott} \emph
  {et~al.}}]{LIGOScientific:2018cki}%
  \BibitemOpen
  \bibfield  {author} {\bibinfo {author} {\bibfnamefont {B.~P.}\ \bibnamefont
  {Abbott}} \emph {et~al.} (\bibinfo {collaboration} {LIGO Scientific,
  Virgo}),\ }\href {\doibase 10.1103/PhysRevLett.121.161101} {\bibfield
  {journal} {\bibinfo  {journal} {Phys. Rev. Lett.}\ }\textbf {\bibinfo
  {volume} {121}},\ \bibinfo {pages} {161101} (\bibinfo {year}
  {2018}{\natexlab{a}})},\ \Eprint {http://arxiv.org/abs/1805.11581}
  {arXiv:1805.11581 [gr-qc]} \BibitemShut {NoStop}%
\bibitem [{\citenamefont {Abbott}\ \emph
  {et~al.}(2021{\natexlab{c}})\citenamefont {Abbott} \emph
  {et~al.}}]{LIGOScientific:2019zcs}%
  \BibitemOpen
  \bibfield  {author} {\bibinfo {author} {\bibfnamefont {B.~P.}\ \bibnamefont
  {Abbott}} \emph {et~al.} (\bibinfo {collaboration} {LIGO Scientific, Virgo,
  VIRGO}),\ }\href {\doibase 10.3847/1538-4357/abdcb7} {\bibfield  {journal}
  {\bibinfo  {journal} {Astrophys. J.}\ }\textbf {\bibinfo {volume} {909}},\
  \bibinfo {pages} {218} (\bibinfo {year} {2021}{\natexlab{c}})},\ \Eprint
  {http://arxiv.org/abs/1908.06060} {arXiv:1908.06060 [astro-ph.CO]}
  \BibitemShut {NoStop}%
\bibitem [{\citenamefont {Nissanke}\ \emph {et~al.}(2010)\citenamefont
  {Nissanke}, \citenamefont {Holz}, \citenamefont {Hughes}, \citenamefont
  {Dalal},\ and\ \citenamefont {Sievers}}]{Nissanke:2009kt}%
  \BibitemOpen
  \bibfield  {author} {\bibinfo {author} {\bibfnamefont {S.}~\bibnamefont
  {Nissanke}}, \bibinfo {author} {\bibfnamefont {D.~E.}\ \bibnamefont {Holz}},
  \bibinfo {author} {\bibfnamefont {S.~A.}\ \bibnamefont {Hughes}}, \bibinfo
  {author} {\bibfnamefont {N.}~\bibnamefont {Dalal}}, \ and\ \bibinfo {author}
  {\bibfnamefont {J.~L.}\ \bibnamefont {Sievers}},\ }\href {\doibase
  10.1088/0004-637X/725/1/496} {\bibfield  {journal} {\bibinfo  {journal}
  {Astrophys. J.}\ }\textbf {\bibinfo {volume} {725}},\ \bibinfo {pages} {496}
  (\bibinfo {year} {2010})},\ \Eprint {http://arxiv.org/abs/0904.1017}
  {arXiv:0904.1017 [astro-ph.CO]} \BibitemShut {NoStop}%
\bibitem [{\citenamefont {Nissanke}\ \emph {et~al.}(2011)\citenamefont
  {Nissanke}, \citenamefont {Sievers}, \citenamefont {Dalal},\ and\
  \citenamefont {Holz}}]{Nissanke:2011ax}%
  \BibitemOpen
  \bibfield  {author} {\bibinfo {author} {\bibfnamefont {S.}~\bibnamefont
  {Nissanke}}, \bibinfo {author} {\bibfnamefont {J.}~\bibnamefont {Sievers}},
  \bibinfo {author} {\bibfnamefont {N.}~\bibnamefont {Dalal}}, \ and\ \bibinfo
  {author} {\bibfnamefont {D.}~\bibnamefont {Holz}},\ }\href {\doibase
  10.1088/0004-637X/739/2/99} {\bibfield  {journal} {\bibinfo  {journal}
  {Astrophys. J.}\ }\textbf {\bibinfo {volume} {739}},\ \bibinfo {pages} {99}
  (\bibinfo {year} {2011})},\ \Eprint {http://arxiv.org/abs/1105.3184}
  {arXiv:1105.3184 [astro-ph.CO]} \BibitemShut {NoStop}%
\bibitem [{\citenamefont {Abbott}\ \emph
  {et~al.}(2021{\natexlab{d}})\citenamefont {Abbott} \emph
  {et~al.}}]{LIGOScientific:2020ibl}%
  \BibitemOpen
  \bibfield  {author} {\bibinfo {author} {\bibfnamefont {R.}~\bibnamefont
  {Abbott}} \emph {et~al.} (\bibinfo {collaboration} {LIGO Scientific,
  Virgo}),\ }\href {\doibase 10.1103/PhysRevX.11.021053} {\bibfield  {journal}
  {\bibinfo  {journal} {Phys. Rev. X}\ }\textbf {\bibinfo {volume} {11}},\
  \bibinfo {pages} {021053} (\bibinfo {year} {2021}{\natexlab{d}})},\ \Eprint
  {http://arxiv.org/abs/2010.14527} {arXiv:2010.14527 [gr-qc]} \BibitemShut
  {NoStop}%
\bibitem [{\citenamefont {Abbott}\ \emph
  {et~al.}(2021{\natexlab{e}})\citenamefont {Abbott} \emph
  {et~al.}}]{LIGOScientific:2021djp}%
  \BibitemOpen
  \bibfield  {author} {\bibinfo {author} {\bibfnamefont {R.}~\bibnamefont
  {Abbott}} \emph {et~al.} (\bibinfo {collaboration} {LIGO Scientific, VIRGO,
  KAGRA}),\ }\href@noop {} {\  (\bibinfo {year} {2021}{\natexlab{e}})},\
  \Eprint {http://arxiv.org/abs/2111.03606} {arXiv:2111.03606 [gr-qc]}
  \BibitemShut {NoStop}%
\bibitem [{\citenamefont {Abbott}\ \emph
  {et~al.}(2021{\natexlab{f}})\citenamefont {Abbott} \emph
  {et~al.}}]{LIGOScientific:2021usb}%
  \BibitemOpen
  \bibfield  {author} {\bibinfo {author} {\bibfnamefont {R.}~\bibnamefont
  {Abbott}} \emph {et~al.} (\bibinfo {collaboration} {LIGO Scientific,
  VIRGO}),\ }\href@noop {} {\  (\bibinfo {year} {2021}{\natexlab{f}})},\
  \Eprint {http://arxiv.org/abs/2108.01045} {arXiv:2108.01045 [gr-qc]}
  \BibitemShut {NoStop}%
\bibitem [{\citenamefont {Nitz}\ \emph {et~al.}(2021)\citenamefont {Nitz},
  \citenamefont {Capano}, \citenamefont {Kumar}, \citenamefont {Wang},
  \citenamefont {Kastha}, \citenamefont {Sch\"afer}, \citenamefont
  {Dhurkunde},\ and\ \citenamefont {Cabero}}]{Nitz:2021uxj}%
  \BibitemOpen
  \bibfield  {author} {\bibinfo {author} {\bibfnamefont {A.~H.}\ \bibnamefont
  {Nitz}}, \bibinfo {author} {\bibfnamefont {C.~D.}\ \bibnamefont {Capano}},
  \bibinfo {author} {\bibfnamefont {S.}~\bibnamefont {Kumar}}, \bibinfo
  {author} {\bibfnamefont {Y.-F.}\ \bibnamefont {Wang}}, \bibinfo {author}
  {\bibfnamefont {S.}~\bibnamefont {Kastha}}, \bibinfo {author} {\bibfnamefont
  {M.}~\bibnamefont {Sch\"afer}}, \bibinfo {author} {\bibfnamefont
  {R.}~\bibnamefont {Dhurkunde}}, \ and\ \bibinfo {author} {\bibfnamefont
  {M.}~\bibnamefont {Cabero}},\ }\href {\doibase 10.3847/1538-4357/ac1c03}
  {\bibfield  {journal} {\bibinfo  {journal} {Astrophys. J.}\ }\textbf
  {\bibinfo {volume} {922}},\ \bibinfo {pages} {76} (\bibinfo {year} {2021})},\
  \Eprint {http://arxiv.org/abs/2105.09151} {arXiv:2105.09151 [astro-ph.HE]}
  \BibitemShut {NoStop}%
\bibitem [{\citenamefont {Venumadhav}\ \emph {et~al.}(2019)\citenamefont
  {Venumadhav}, \citenamefont {Zackay}, \citenamefont {Roulet}, \citenamefont
  {Dai},\ and\ \citenamefont {Zaldarriaga}}]{Venumadhav:2019aa}%
  \BibitemOpen
  \bibfield  {author} {\bibinfo {author} {\bibfnamefont {T.}~\bibnamefont
  {Venumadhav}}, \bibinfo {author} {\bibfnamefont {B.}~\bibnamefont {Zackay}},
  \bibinfo {author} {\bibfnamefont {J.}~\bibnamefont {Roulet}}, \bibinfo
  {author} {\bibfnamefont {L.}~\bibnamefont {Dai}}, \ and\ \bibinfo {author}
  {\bibfnamefont {M.}~\bibnamefont {Zaldarriaga}},\ }\href {\doibase
  10.1103/PhysRevD.100.023011} {\bibfield  {journal} {\bibinfo  {journal}
  {Phys. Rev. D}\ }\textbf {\bibinfo {volume} {100}},\ \bibinfo {pages}
  {023011} (\bibinfo {year} {2019})}\BibitemShut {NoStop}%
\bibitem [{\citenamefont {Venumadhav}\ \emph {et~al.}(2020)\citenamefont
  {Venumadhav}, \citenamefont {Zackay}, \citenamefont {Roulet}, \citenamefont
  {Dai},\ and\ \citenamefont {Zaldarriaga}}]{Venumadhav:2020aa}%
  \BibitemOpen
  \bibfield  {author} {\bibinfo {author} {\bibfnamefont {T.}~\bibnamefont
  {Venumadhav}}, \bibinfo {author} {\bibfnamefont {B.}~\bibnamefont {Zackay}},
  \bibinfo {author} {\bibfnamefont {J.}~\bibnamefont {Roulet}}, \bibinfo
  {author} {\bibfnamefont {L.}~\bibnamefont {Dai}}, \ and\ \bibinfo {author}
  {\bibfnamefont {M.}~\bibnamefont {Zaldarriaga}},\ }\href {\doibase
  10.1103/PhysRevD.101.083030} {\bibfield  {journal} {\bibinfo  {journal}
  {Phys. Rev. D}\ }\textbf {\bibinfo {volume} {101}},\ \bibinfo {pages}
  {083030} (\bibinfo {year} {2020})}\BibitemShut {NoStop}%
\bibitem [{\citenamefont {Abbott}\ \emph
  {et~al.}(2018{\natexlab{b}})\citenamefont {Abbott} \emph
  {et~al.}}]{KAGRA:2013rdx}%
  \BibitemOpen
  \bibfield  {author} {\bibinfo {author} {\bibfnamefont {B.~P.}\ \bibnamefont
  {Abbott}} \emph {et~al.} (\bibinfo {collaboration} {KAGRA, LIGO Scientific,
  Virgo, VIRGO}),\ }\href {\doibase 10.1007/s41114-020-00026-9} {\bibfield
  {journal} {\bibinfo  {journal} {Living Rev. Rel.}\ }\textbf {\bibinfo
  {volume} {21}},\ \bibinfo {pages} {3} (\bibinfo {year}
  {2018}{\natexlab{b}})},\ \Eprint {http://arxiv.org/abs/1304.0670}
  {arXiv:1304.0670 [gr-qc]} \BibitemShut {NoStop}%
\bibitem [{\citenamefont {Akutsu}\ \emph {et~al.}(2021)\citenamefont {Akutsu}
  \emph {et~al.}}]{KAGRA:2020tym}%
  \BibitemOpen
  \bibfield  {author} {\bibinfo {author} {\bibfnamefont {T.}~\bibnamefont
  {Akutsu}} \emph {et~al.} (\bibinfo {collaboration} {KAGRA}),\ }\href
  {\doibase 10.1093/ptep/ptaa125} {\bibfield  {journal} {\bibinfo  {journal}
  {PTEP}\ }\textbf {\bibinfo {volume} {2021}},\ \bibinfo {pages} {05A101}
  (\bibinfo {year} {2021})},\ \Eprint {http://arxiv.org/abs/2005.05574}
  {arXiv:2005.05574 [physics.ins-det]} \BibitemShut {NoStop}%
\bibitem [{\citenamefont {Iyer}\ \emph {et~al.}(2021)\citenamefont {Iyer} \emph
  {et~al.}}]{IndiGO:2011aaa}%
  \BibitemOpen
  \bibfield  {author} {\bibinfo {author} {\bibfnamefont {B.}~\bibnamefont
  {Iyer}} \emph {et~al.},\ }\href
  {https://dcc.ligo.org/public/0075/M1100296/002/LIGO-India_lw-v2.pdf}
  {\bibfield  {journal} {\bibinfo  {journal} {{LIGO-India: Proposal for an
  interferometric gravitational wave observatory.}}\ } (\bibinfo {year}
  {2021})}\BibitemShut {NoStop}%
\bibitem [{\citenamefont {Unnikrishnan}(2013)}]{Unnikrishnan:2013qwa}%
  \BibitemOpen
  \bibfield  {author} {\bibinfo {author} {\bibfnamefont {C.~S.}\ \bibnamefont
  {Unnikrishnan}},\ }\href {\doibase 10.1142/S0218271813410101} {\bibfield
  {journal} {\bibinfo  {journal} {Int. J. Mod. Phys. D}\ }\textbf {\bibinfo
  {volume} {22}},\ \bibinfo {pages} {1341010} (\bibinfo {year} {2013})},\
  \Eprint {http://arxiv.org/abs/1510.06059} {arXiv:1510.06059
  [physics.ins-det]} \BibitemShut {NoStop}%
\bibitem [{\citenamefont {Adhikari}\ \emph {et~al.}(2020)\citenamefont
  {Adhikari} \emph {et~al.}}]{LIGO:2020xsf}%
  \BibitemOpen
  \bibfield  {author} {\bibinfo {author} {\bibfnamefont {R.~X.}\ \bibnamefont
  {Adhikari}} \emph {et~al.} (\bibinfo {collaboration} {LIGO}),\ }\href
  {\doibase 10.1088/1361-6382/ab9143} {\bibfield  {journal} {\bibinfo
  {journal} {Class. Quant. Grav.}\ }\textbf {\bibinfo {volume} {37}},\ \bibinfo
  {pages} {165003} (\bibinfo {year} {2020})},\ \Eprint
  {http://arxiv.org/abs/2001.11173} {arXiv:2001.11173 [astro-ph.IM]}
  \BibitemShut {NoStop}%
\bibitem [{\citenamefont {Maggiore}\ \emph {et~al.}(2020)\citenamefont
  {Maggiore} \emph {et~al.}}]{Maggiore:2019uih}%
  \BibitemOpen
  \bibfield  {author} {\bibinfo {author} {\bibfnamefont {M.}~\bibnamefont
  {Maggiore}} \emph {et~al.},\ }\href {\doibase 10.1088/1475-7516/2020/03/050}
  {\bibfield  {journal} {\bibinfo  {journal} {JCAP}\ }\textbf {\bibinfo
  {volume} {03}},\ \bibinfo {pages} {050} (\bibinfo {year} {2020})},\ \Eprint
  {http://arxiv.org/abs/1912.02622} {arXiv:1912.02622 [astro-ph.CO]}
  \BibitemShut {NoStop}%
\bibitem [{\citenamefont {Reitze}\ \emph {et~al.}(2019)\citenamefont {Reitze}
  \emph {et~al.}}]{Reitze:2019iox}%
  \BibitemOpen
  \bibfield  {author} {\bibinfo {author} {\bibfnamefont {D.}~\bibnamefont
  {Reitze}} \emph {et~al.},\ }\href@noop {} {\bibfield  {journal} {\bibinfo
  {journal} {Bull. Am. Astron. Soc.}\ }\textbf {\bibinfo {volume} {51}},\
  \bibinfo {pages} {035} (\bibinfo {year} {2019})},\ \Eprint
  {http://arxiv.org/abs/1907.04833} {arXiv:1907.04833 [astro-ph.IM]}
  \BibitemShut {NoStop}%
\bibitem [{\citenamefont {Amaro-Seoane}\ \emph {et~al.}(2017)\citenamefont
  {Amaro-Seoane}, \citenamefont {Audley}, \citenamefont {Babak}, \citenamefont
  {Baker}, \citenamefont {Barausse}, \citenamefont {Bender}, \citenamefont
  {Berti} \emph {et~al.}}]{LISA:2017aaa}%
  \BibitemOpen
  \bibfield  {author} {\bibinfo {author} {\bibfnamefont {P.}~\bibnamefont
  {Amaro-Seoane}}, \bibinfo {author} {\bibfnamefont {H.}~\bibnamefont
  {Audley}}, \bibinfo {author} {\bibfnamefont {S.}~\bibnamefont {Babak}},
  \bibinfo {author} {\bibfnamefont {J.}~\bibnamefont {Baker}}, \bibinfo
  {author} {\bibfnamefont {E.}~\bibnamefont {Barausse}}, \bibinfo {author}
  {\bibfnamefont {P.}~\bibnamefont {Bender}}, \bibinfo {author} {\bibfnamefont
  {E.}~\bibnamefont {Berti}},  \emph {et~al.},\ }\href@noop {} {\  (\bibinfo
  {year} {2017})},\ \Eprint {http://arxiv.org/abs/1702.00786} {arXiv:1702.00786
  [astro-ph.IM]} \BibitemShut {NoStop}%
\bibitem [{\citenamefont {Littenberg}\ and\ \citenamefont
  {Cornish}(2023)}]{Littenberg:2023xpl}%
  \BibitemOpen
  \bibfield  {author} {\bibinfo {author} {\bibfnamefont {T.~B.}\ \bibnamefont
  {Littenberg}}\ and\ \bibinfo {author} {\bibfnamefont {N.~J.}\ \bibnamefont
  {Cornish}},\ }\href {\doibase 10.1103/PhysRevD.107.063004} {\bibfield
  {journal} {\bibinfo  {journal} {Phys. Rev. D}\ }\textbf {\bibinfo {volume}
  {107}},\ \bibinfo {pages} {063004} (\bibinfo {year} {2023})},\ \Eprint
  {http://arxiv.org/abs/2301.03673} {arXiv:2301.03673 [gr-qc]} \BibitemShut
  {NoStop}%
\bibitem [{\citenamefont {Babak}\ \emph {et~al.}(2017)\citenamefont {Babak},
  \citenamefont {Gair}, \citenamefont {Sesana}, \citenamefont {Barausse},
  \citenamefont {Sopuerta}, \citenamefont {Berry}, \citenamefont {Berti},
  \citenamefont {Amaro-Seoane}, \citenamefont {Petiteau},\ and\ \citenamefont
  {Klein}}]{Babak:2017tow}%
  \BibitemOpen
  \bibfield  {author} {\bibinfo {author} {\bibfnamefont {S.}~\bibnamefont
  {Babak}}, \bibinfo {author} {\bibfnamefont {J.}~\bibnamefont {Gair}},
  \bibinfo {author} {\bibfnamefont {A.}~\bibnamefont {Sesana}}, \bibinfo
  {author} {\bibfnamefont {E.}~\bibnamefont {Barausse}}, \bibinfo {author}
  {\bibfnamefont {C.~F.}\ \bibnamefont {Sopuerta}}, \bibinfo {author}
  {\bibfnamefont {C.~P.~L.}\ \bibnamefont {Berry}}, \bibinfo {author}
  {\bibfnamefont {E.}~\bibnamefont {Berti}}, \bibinfo {author} {\bibfnamefont
  {P.}~\bibnamefont {Amaro-Seoane}}, \bibinfo {author} {\bibfnamefont
  {A.}~\bibnamefont {Petiteau}}, \ and\ \bibinfo {author} {\bibfnamefont
  {A.}~\bibnamefont {Klein}},\ }\href {\doibase 10.1103/PhysRevD.95.103012}
  {\bibfield  {journal} {\bibinfo  {journal} {Phys. Rev. D}\ }\textbf {\bibinfo
  {volume} {95}},\ \bibinfo {pages} {103012} (\bibinfo {year} {2017})},\
  \Eprint {http://arxiv.org/abs/1703.09722} {arXiv:1703.09722 [gr-qc]}
  \BibitemShut {NoStop}%
\bibitem [{\citenamefont {Spadaro}\ \emph {et~al.}(2023)\citenamefont
  {Spadaro}, \citenamefont {Buscicchio}, \citenamefont {Vetrugno},
  \citenamefont {Klein}, \citenamefont {Gerosa}, \citenamefont {Vitale},
  \citenamefont {Dolesi}, \citenamefont {Weber},\ and\ \citenamefont
  {Colpi}}]{Spadaro:2023muy}%
  \BibitemOpen
  \bibfield  {author} {\bibinfo {author} {\bibfnamefont {A.}~\bibnamefont
  {Spadaro}}, \bibinfo {author} {\bibfnamefont {R.}~\bibnamefont {Buscicchio}},
  \bibinfo {author} {\bibfnamefont {D.}~\bibnamefont {Vetrugno}}, \bibinfo
  {author} {\bibfnamefont {A.}~\bibnamefont {Klein}}, \bibinfo {author}
  {\bibfnamefont {D.}~\bibnamefont {Gerosa}}, \bibinfo {author} {\bibfnamefont
  {S.}~\bibnamefont {Vitale}}, \bibinfo {author} {\bibfnamefont
  {R.}~\bibnamefont {Dolesi}}, \bibinfo {author} {\bibfnamefont {W.~J.}\
  \bibnamefont {Weber}}, \ and\ \bibinfo {author} {\bibfnamefont
  {M.}~\bibnamefont {Colpi}},\ }\href@noop {} {\  (\bibinfo {year} {2023})},\
  \Eprint {http://arxiv.org/abs/2306.03923} {arXiv:2306.03923 [gr-qc]}
  \BibitemShut {NoStop}%
\bibitem [{\citenamefont {Weaving}\ \emph {et~al.}(2023)\citenamefont
  {Weaving}, \citenamefont {Nuttall}, \citenamefont {Harry}, \citenamefont
  {Wu},\ and\ \citenamefont {Nitz}}]{Weaving:2023fji}%
  \BibitemOpen
  \bibfield  {author} {\bibinfo {author} {\bibfnamefont {C.~R.}\ \bibnamefont
  {Weaving}}, \bibinfo {author} {\bibfnamefont {L.~K.}\ \bibnamefont
  {Nuttall}}, \bibinfo {author} {\bibfnamefont {I.~W.}\ \bibnamefont {Harry}},
  \bibinfo {author} {\bibfnamefont {S.}~\bibnamefont {Wu}}, \ and\ \bibinfo
  {author} {\bibfnamefont {A.}~\bibnamefont {Nitz}},\ }\href@noop {} {\
  (\bibinfo {year} {2023})},\ \Eprint {http://arxiv.org/abs/2306.16429}
  {arXiv:2306.16429 [astro-ph.IM]} \BibitemShut {NoStop}%
\bibitem [{\citenamefont {Sah}\ and\ \citenamefont
  {Mukherjee}(2023)}]{Sah:2023bgr}%
  \BibitemOpen
  \bibfield  {author} {\bibinfo {author} {\bibfnamefont {M.~R.}\ \bibnamefont
  {Sah}}\ and\ \bibinfo {author} {\bibfnamefont {S.}~\bibnamefont
  {Mukherjee}},\ }\href@noop {} {\  (\bibinfo {year} {2023})},\ \Eprint
  {http://arxiv.org/abs/2307.06405} {arXiv:2307.06405 [gr-qc]} \BibitemShut
  {NoStop}%
\bibitem [{\citenamefont {Pozzoli}\ \emph {et~al.}(2023)\citenamefont
  {Pozzoli}, \citenamefont {Babak}, \citenamefont {Sesana}, \citenamefont
  {Bonetti},\ and\ \citenamefont {Karnesis}}]{Pozzoli:2023kxy}%
  \BibitemOpen
  \bibfield  {author} {\bibinfo {author} {\bibfnamefont {F.}~\bibnamefont
  {Pozzoli}}, \bibinfo {author} {\bibfnamefont {S.}~\bibnamefont {Babak}},
  \bibinfo {author} {\bibfnamefont {A.}~\bibnamefont {Sesana}}, \bibinfo
  {author} {\bibfnamefont {M.}~\bibnamefont {Bonetti}}, \ and\ \bibinfo
  {author} {\bibfnamefont {N.}~\bibnamefont {Karnesis}},\ }\href@noop {} {\
  (\bibinfo {year} {2023})},\ \Eprint {http://arxiv.org/abs/2302.07043}
  {arXiv:2302.07043 [astro-ph.GA]} \BibitemShut {NoStop}%
\bibitem [{\citenamefont {Flauger}\ \emph {et~al.}(2021)\citenamefont
  {Flauger}, \citenamefont {Karnesis}, \citenamefont {Nardini}, \citenamefont
  {Pieroni}, \citenamefont {Ricciardone},\ and\ \citenamefont
  {Torrado}}]{Flauger:2020qyi}%
  \BibitemOpen
  \bibfield  {author} {\bibinfo {author} {\bibfnamefont {R.}~\bibnamefont
  {Flauger}}, \bibinfo {author} {\bibfnamefont {N.}~\bibnamefont {Karnesis}},
  \bibinfo {author} {\bibfnamefont {G.}~\bibnamefont {Nardini}}, \bibinfo
  {author} {\bibfnamefont {M.}~\bibnamefont {Pieroni}}, \bibinfo {author}
  {\bibfnamefont {A.}~\bibnamefont {Ricciardone}}, \ and\ \bibinfo {author}
  {\bibfnamefont {J.}~\bibnamefont {Torrado}},\ }\href {\doibase
  10.1088/1475-7516/2021/01/059} {\bibfield  {journal} {\bibinfo  {journal}
  {JCAP}\ }\textbf {\bibinfo {volume} {01}},\ \bibinfo {pages} {059} (\bibinfo
  {year} {2021})},\ \Eprint {http://arxiv.org/abs/2009.11845} {arXiv:2009.11845
  [astro-ph.CO]} \BibitemShut {NoStop}%
\bibitem [{\citenamefont {Bandopadhyay}\ and\ \citenamefont
  {Moore}(2023)}]{Bandopadhyay:2023gkb}%
  \BibitemOpen
  \bibfield  {author} {\bibinfo {author} {\bibfnamefont {D.}~\bibnamefont
  {Bandopadhyay}}\ and\ \bibinfo {author} {\bibfnamefont {C.~J.}\ \bibnamefont
  {Moore}},\ }\href@noop {} {\  (\bibinfo {year} {2023})},\ \Eprint
  {http://arxiv.org/abs/2305.18048} {arXiv:2305.18048 [gr-qc]} \BibitemShut
  {NoStop}%
\bibitem [{\citenamefont {Katz}\ \emph {et~al.}(2021)\citenamefont {Katz},
  \citenamefont {Chua}, \citenamefont {Speri}, \citenamefont {Warburton},\ and\
  \citenamefont {Hughes}}]{Katz:2021yft}%
  \BibitemOpen
  \bibfield  {author} {\bibinfo {author} {\bibfnamefont {M.~L.}\ \bibnamefont
  {Katz}}, \bibinfo {author} {\bibfnamefont {A.~J.~K.}\ \bibnamefont {Chua}},
  \bibinfo {author} {\bibfnamefont {L.}~\bibnamefont {Speri}}, \bibinfo
  {author} {\bibfnamefont {N.}~\bibnamefont {Warburton}}, \ and\ \bibinfo
  {author} {\bibfnamefont {S.~A.}\ \bibnamefont {Hughes}},\ }\href {\doibase
  10.1103/PhysRevD.104.064047} {\bibfield  {journal} {\bibinfo  {journal}
  {Phys. Rev. D}\ }\textbf {\bibinfo {volume} {104}},\ \bibinfo {pages}
  {064047} (\bibinfo {year} {2021})},\ \Eprint
  {http://arxiv.org/abs/2104.04582} {arXiv:2104.04582 [gr-qc]} \BibitemShut
  {NoStop}%
\bibitem [{\citenamefont {Pizzati}\ \emph {et~al.}(2022)\citenamefont
  {Pizzati}, \citenamefont {Sachdev}, \citenamefont {Gupta},\ and\
  \citenamefont {Sathyaprakash}}]{Pizzati:2021apa}%
  \BibitemOpen
  \bibfield  {author} {\bibinfo {author} {\bibfnamefont {E.}~\bibnamefont
  {Pizzati}}, \bibinfo {author} {\bibfnamefont {S.}~\bibnamefont {Sachdev}},
  \bibinfo {author} {\bibfnamefont {A.}~\bibnamefont {Gupta}}, \ and\ \bibinfo
  {author} {\bibfnamefont {B.}~\bibnamefont {Sathyaprakash}},\ }\href {\doibase
  10.1103/PhysRevD.105.104016} {\bibfield  {journal} {\bibinfo  {journal}
  {Phys. Rev. D}\ }\textbf {\bibinfo {volume} {105}},\ \bibinfo {pages}
  {104016} (\bibinfo {year} {2022})},\ \Eprint
  {http://arxiv.org/abs/2102.07692} {arXiv:2102.07692 [gr-qc]} \BibitemShut
  {NoStop}%
\bibitem [{\citenamefont {Antonelli}\ \emph {et~al.}(2021)\citenamefont
  {Antonelli}, \citenamefont {Burke},\ and\ \citenamefont
  {Gair}}]{Antonelli:2021vwg}%
  \BibitemOpen
  \bibfield  {author} {\bibinfo {author} {\bibfnamefont {A.}~\bibnamefont
  {Antonelli}}, \bibinfo {author} {\bibfnamefont {O.}~\bibnamefont {Burke}}, \
  and\ \bibinfo {author} {\bibfnamefont {J.~R.}\ \bibnamefont {Gair}},\ }\href
  {\doibase 10.1093/mnras/stab2358} {\bibfield  {journal} {\bibinfo  {journal}
  {Mon. Not. Roy. Astron. Soc.}\ }\textbf {\bibinfo {volume} {507}},\ \bibinfo
  {pages} {5069} (\bibinfo {year} {2021})},\ \Eprint
  {http://arxiv.org/abs/2104.01897} {arXiv:2104.01897 [gr-qc]} \BibitemShut
  {NoStop}%
\bibitem [{\citenamefont {Samajdar}\ \emph {et~al.}(2021)\citenamefont
  {Samajdar}, \citenamefont {Janquart}, \citenamefont {Van Den~Broeck},\ and\
  \citenamefont {Dietrich}}]{Samajdar:2021egv}%
  \BibitemOpen
  \bibfield  {author} {\bibinfo {author} {\bibfnamefont {A.}~\bibnamefont
  {Samajdar}}, \bibinfo {author} {\bibfnamefont {J.}~\bibnamefont {Janquart}},
  \bibinfo {author} {\bibfnamefont {C.}~\bibnamefont {Van Den~Broeck}}, \ and\
  \bibinfo {author} {\bibfnamefont {T.}~\bibnamefont {Dietrich}},\ }\href
  {\doibase 10.1103/PhysRevD.104.044003} {\bibfield  {journal} {\bibinfo
  {journal} {Phys. Rev. D}\ }\textbf {\bibinfo {volume} {104}},\ \bibinfo
  {pages} {044003} (\bibinfo {year} {2021})},\ \Eprint
  {http://arxiv.org/abs/2102.07544} {arXiv:2102.07544 [gr-qc]} \BibitemShut
  {NoStop}%
\bibitem [{\citenamefont {Janquart}\ \emph {et~al.}(2022)\citenamefont
  {Janquart}, \citenamefont {Baka}, \citenamefont {Samajdar}, \citenamefont
  {Dietrich},\ and\ \citenamefont {Van Den~Broeck}}]{Janquart:2022nyz}%
  \BibitemOpen
  \bibfield  {author} {\bibinfo {author} {\bibfnamefont {J.}~\bibnamefont
  {Janquart}}, \bibinfo {author} {\bibfnamefont {T.}~\bibnamefont {Baka}},
  \bibinfo {author} {\bibfnamefont {A.}~\bibnamefont {Samajdar}}, \bibinfo
  {author} {\bibfnamefont {T.}~\bibnamefont {Dietrich}}, \ and\ \bibinfo
  {author} {\bibfnamefont {C.}~\bibnamefont {Van Den~Broeck}},\ }\href@noop {}
  {\  (\bibinfo {year} {2022})},\ \Eprint {http://arxiv.org/abs/2211.01304}
  {arXiv:2211.01304 [gr-qc]} \BibitemShut {NoStop}%
\bibitem [{\citenamefont {Langendorff}\ \emph {et~al.}(2023)\citenamefont
  {Langendorff}, \citenamefont {Kolmus}, \citenamefont {Janquart},\ and\
  \citenamefont {Van Den~Broeck}}]{Langendorff:2022fzq}%
  \BibitemOpen
  \bibfield  {author} {\bibinfo {author} {\bibfnamefont {J.}~\bibnamefont
  {Langendorff}}, \bibinfo {author} {\bibfnamefont {A.}~\bibnamefont {Kolmus}},
  \bibinfo {author} {\bibfnamefont {J.}~\bibnamefont {Janquart}}, \ and\
  \bibinfo {author} {\bibfnamefont {C.}~\bibnamefont {Van Den~Broeck}},\ }\href
  {\doibase 10.1103/PhysRevLett.130.171402} {\bibfield  {journal} {\bibinfo
  {journal} {Phys. Rev. Lett.}\ }\textbf {\bibinfo {volume} {130}},\ \bibinfo
  {pages} {171402} (\bibinfo {year} {2023})},\ \Eprint
  {http://arxiv.org/abs/2211.15097} {arXiv:2211.15097 [gr-qc]} \BibitemShut
  {NoStop}%
\bibitem [{\citenamefont {Relton}\ and\ \citenamefont
  {Raymond}(2021)}]{Relton:2021cax}%
  \BibitemOpen
  \bibfield  {author} {\bibinfo {author} {\bibfnamefont {P.}~\bibnamefont
  {Relton}}\ and\ \bibinfo {author} {\bibfnamefont {V.}~\bibnamefont
  {Raymond}},\ }\href {\doibase 10.1103/PhysRevD.104.084039} {\bibfield
  {journal} {\bibinfo  {journal} {Phys. Rev. D}\ }\textbf {\bibinfo {volume}
  {104}},\ \bibinfo {pages} {084039} (\bibinfo {year} {2021})},\ \Eprint
  {http://arxiv.org/abs/2103.16225} {arXiv:2103.16225 [gr-qc]} \BibitemShut
  {NoStop}%
\bibitem [{\citenamefont {Relton}\ \emph {et~al.}(2022)\citenamefont {Relton},
  \citenamefont {Virtuoso}, \citenamefont {Bini}, \citenamefont {Raymond},
  \citenamefont {Harry}, \citenamefont {Drago}, \citenamefont {Lazzaro},
  \citenamefont {Miani},\ and\ \citenamefont {Tiwari}}]{Relton:2022whr}%
  \BibitemOpen
  \bibfield  {author} {\bibinfo {author} {\bibfnamefont {P.}~\bibnamefont
  {Relton}}, \bibinfo {author} {\bibfnamefont {A.}~\bibnamefont {Virtuoso}},
  \bibinfo {author} {\bibfnamefont {S.}~\bibnamefont {Bini}}, \bibinfo {author}
  {\bibfnamefont {V.}~\bibnamefont {Raymond}}, \bibinfo {author} {\bibfnamefont
  {I.}~\bibnamefont {Harry}}, \bibinfo {author} {\bibfnamefont
  {M.}~\bibnamefont {Drago}}, \bibinfo {author} {\bibfnamefont
  {C.}~\bibnamefont {Lazzaro}}, \bibinfo {author} {\bibfnamefont
  {A.}~\bibnamefont {Miani}}, \ and\ \bibinfo {author} {\bibfnamefont
  {S.}~\bibnamefont {Tiwari}},\ }\href {\doibase 10.1103/PhysRevD.106.104045}
  {\bibfield  {journal} {\bibinfo  {journal} {Phys. Rev. D}\ }\textbf {\bibinfo
  {volume} {106}},\ \bibinfo {pages} {104045} (\bibinfo {year} {2022})},\
  \Eprint {http://arxiv.org/abs/2208.00261} {arXiv:2208.00261 [gr-qc]}
  \BibitemShut {NoStop}%
\bibitem [{\citenamefont {Ashton}\ \emph {et~al.}(2019)\citenamefont {Ashton}
  \emph {et~al.}}]{Ashton:2018jfp}%
  \BibitemOpen
  \bibfield  {author} {\bibinfo {author} {\bibfnamefont {G.}~\bibnamefont
  {Ashton}} \emph {et~al.},\ }\href {\doibase 10.3847/1538-4365/ab06fc}
  {\bibfield  {journal} {\bibinfo  {journal} {Astrophys. J. Suppl.}\ }\textbf
  {\bibinfo {volume} {241}},\ \bibinfo {pages} {27} (\bibinfo {year} {2019})},\
  \Eprint {http://arxiv.org/abs/1811.02042} {arXiv:1811.02042 [astro-ph.IM]}
  \BibitemShut {NoStop}%
\bibitem [{\citenamefont {Romero-Shaw}\ \emph {et~al.}(2020)\citenamefont
  {Romero-Shaw} \emph {et~al.}}]{Romero-Shaw:2020owr}%
  \BibitemOpen
  \bibfield  {author} {\bibinfo {author} {\bibfnamefont {I.~M.}\ \bibnamefont
  {Romero-Shaw}} \emph {et~al.},\ }\href {\doibase 10.1093/mnras/staa2850}
  {\bibfield  {journal} {\bibinfo  {journal} {Mon. Not. Roy. Astron. Soc.}\
  }\textbf {\bibinfo {volume} {499}},\ \bibinfo {pages} {3295} (\bibinfo {year}
  {2020})},\ \Eprint {http://arxiv.org/abs/2006.00714} {arXiv:2006.00714
  [astro-ph.IM]} \BibitemShut {NoStop}%
\bibitem [{\citenamefont {Biwer}\ \emph {et~al.}(2019)\citenamefont {Biwer},
  \citenamefont {Capano}, \citenamefont {De}, \citenamefont {Cabero},
  \citenamefont {Brown}, \citenamefont {Nitz},\ and\ \citenamefont
  {Raymond}}]{Biwer:2018osg}%
  \BibitemOpen
  \bibfield  {author} {\bibinfo {author} {\bibfnamefont {C.~M.}\ \bibnamefont
  {Biwer}}, \bibinfo {author} {\bibfnamefont {C.~D.}\ \bibnamefont {Capano}},
  \bibinfo {author} {\bibfnamefont {S.}~\bibnamefont {De}}, \bibinfo {author}
  {\bibfnamefont {M.}~\bibnamefont {Cabero}}, \bibinfo {author} {\bibfnamefont
  {D.~A.}\ \bibnamefont {Brown}}, \bibinfo {author} {\bibfnamefont {A.~H.}\
  \bibnamefont {Nitz}}, \ and\ \bibinfo {author} {\bibfnamefont
  {V.}~\bibnamefont {Raymond}},\ }\href {\doibase 10.1088/1538-3873/aaef0b}
  {\bibfield  {journal} {\bibinfo  {journal} {Publ. Astron. Soc. Pac.}\
  }\textbf {\bibinfo {volume} {131}},\ \bibinfo {pages} {024503} (\bibinfo
  {year} {2019})},\ \Eprint {http://arxiv.org/abs/1807.10312} {arXiv:1807.10312
  [astro-ph.IM]} \BibitemShut {NoStop}%
\bibitem [{\citenamefont {Veitch}\ \emph {et~al.}(2015)\citenamefont {Veitch}
  \emph {et~al.}}]{Veitch:2014wba}%
  \BibitemOpen
  \bibfield  {author} {\bibinfo {author} {\bibfnamefont {J.}~\bibnamefont
  {Veitch}} \emph {et~al.},\ }\href {\doibase 10.1103/PhysRevD.91.042003}
  {\bibfield  {journal} {\bibinfo  {journal} {Phys. Rev. D}\ }\textbf {\bibinfo
  {volume} {91}},\ \bibinfo {pages} {042003} (\bibinfo {year} {2015})},\
  \Eprint {http://arxiv.org/abs/1409.7215} {arXiv:1409.7215 [gr-qc]}
  \BibitemShut {NoStop}%
\bibitem [{\citenamefont {Foreman-Mackey}\ \emph {et~al.}(2013)\citenamefont
  {Foreman-Mackey}, \citenamefont {Hogg}, \citenamefont {Lang},\ and\
  \citenamefont {Goodman}}]{Foreman-Mackey:2012any}%
  \BibitemOpen
  \bibfield  {author} {\bibinfo {author} {\bibfnamefont {D.}~\bibnamefont
  {Foreman-Mackey}}, \bibinfo {author} {\bibfnamefont {D.~W.}\ \bibnamefont
  {Hogg}}, \bibinfo {author} {\bibfnamefont {D.}~\bibnamefont {Lang}}, \ and\
  \bibinfo {author} {\bibfnamefont {J.}~\bibnamefont {Goodman}},\ }\href
  {\doibase 10.1086/670067} {\bibfield  {journal} {\bibinfo  {journal} {Publ.
  Astron. Soc. Pac.}\ }\textbf {\bibinfo {volume} {125}},\ \bibinfo {pages}
  {306} (\bibinfo {year} {2013})},\ \Eprint {http://arxiv.org/abs/1202.3665}
  {arXiv:1202.3665 [astro-ph.IM]} \BibitemShut {NoStop}%
\bibitem [{\citenamefont {{Vousden}}\ \emph {et~al.}(2016)\citenamefont
  {{Vousden}}, \citenamefont {{Farr}},\ and\ \citenamefont
  {{Mandel}}}]{Vousden:2016aaa}%
  \BibitemOpen
  \bibfield  {author} {\bibinfo {author} {\bibfnamefont {W.~D.}\ \bibnamefont
  {{Vousden}}}, \bibinfo {author} {\bibfnamefont {W.~M.}\ \bibnamefont
  {{Farr}}}, \ and\ \bibinfo {author} {\bibfnamefont {I.}~\bibnamefont
  {{Mandel}}},\ }\href {\doibase 10.1093/mnras/stv2422} {\bibfield  {journal}
  {\bibinfo  {journal} {Mon. Not. Roy. Astron. Soc.}\ }\textbf {\bibinfo
  {volume} {455}},\ \bibinfo {pages} {1919} (\bibinfo {year} {2016})},\ \Eprint
  {http://arxiv.org/abs/1501.05823} {arXiv:1501.05823} \BibitemShut {NoStop}%
\bibitem [{\citenamefont {Skilling}(2006)}]{Skilling:2006gxv}%
  \BibitemOpen
  \bibfield  {author} {\bibinfo {author} {\bibfnamefont {J.}~\bibnamefont
  {Skilling}},\ }\href {\doibase 10.1214/06-BA127} {\bibfield  {journal}
  {\bibinfo  {journal} {Bayesian Analysis}\ }\textbf {\bibinfo {volume} {1}},\
  \bibinfo {pages} {833} (\bibinfo {year} {2006})}\BibitemShut {NoStop}%
\bibitem [{\citenamefont {{Handley}}\ \emph {et~al.}(2015)\citenamefont
  {{Handley}}, \citenamefont {{Hobson}},\ and\ \citenamefont
  {{Lasenby}}}]{Handley:2015aaa}%
  \BibitemOpen
  \bibfield  {author} {\bibinfo {author} {\bibfnamefont {W.~J.}\ \bibnamefont
  {{Handley}}}, \bibinfo {author} {\bibfnamefont {M.~P.}\ \bibnamefont
  {{Hobson}}}, \ and\ \bibinfo {author} {\bibfnamefont {A.~N.}\ \bibnamefont
  {{Lasenby}}},\ }\href {\doibase 10.1093/mnras/stv1911} {\bibfield  {journal}
  {\bibinfo  {journal} {Mon. Not. Roy. Astron. Soc.}\ }\textbf {\bibinfo
  {volume} {453}},\ \bibinfo {pages} {4384} (\bibinfo {year} {2015})},\ \Eprint
  {http://arxiv.org/abs/1506.00171} {arXiv:1506.00171 [astro-ph.IM]}
  \BibitemShut {NoStop}%
\bibitem [{\citenamefont {Ashton}\ \emph {et~al.}(2022)\citenamefont {Ashton}
  \emph {et~al.}}]{Ashton:2022grj}%
  \BibitemOpen
  \bibfield  {author} {\bibinfo {author} {\bibfnamefont {G.}~\bibnamefont
  {Ashton}} \emph {et~al.},\ }\href {\doibase 10.1038/s43586-022-00121-x}
  {\bibfield  {journal} {\bibinfo  {journal} {Nature}\ }\textbf {\bibinfo
  {volume} {2}} (\bibinfo {year} {2022}),\ 10.1038/s43586-022-00121-x},\
  \Eprint {http://arxiv.org/abs/2205.15570} {arXiv:2205.15570 [stat.CO]}
  \BibitemShut {NoStop}%
\bibitem [{\citenamefont {{Speagle}}(2020)}]{Speagle:2020aaa}%
  \BibitemOpen
  \bibfield  {author} {\bibinfo {author} {\bibfnamefont {J.~S.}\ \bibnamefont
  {{Speagle}}},\ }\href {\doibase 10.1093/mnras/staa278} {\bibfield  {journal}
  {\bibinfo  {journal} {Mon. Not. Roy. Astron. Soc.}\ }\textbf {\bibinfo
  {volume} {493}},\ \bibinfo {pages} {3132} (\bibinfo {year} {2020})},\ \Eprint
  {http://arxiv.org/abs/1904.02180} {arXiv:1904.02180 [astro-ph.IM]}
  \BibitemShut {NoStop}%
\bibitem [{\citenamefont {Edwards}\ \emph {et~al.}(2023)\citenamefont
  {Edwards}, \citenamefont {Wong}, \citenamefont {Lam}, \citenamefont {Coogan},
  \citenamefont {Foreman-Mackey}, \citenamefont {Isi},\ and\ \citenamefont
  {Zimmerman}}]{Edwards:2023sak}%
  \BibitemOpen
  \bibfield  {author} {\bibinfo {author} {\bibfnamefont {T.~D.~P.}\
  \bibnamefont {Edwards}}, \bibinfo {author} {\bibfnamefont {K.~W.~K.}\
  \bibnamefont {Wong}}, \bibinfo {author} {\bibfnamefont {K.~K.~H.}\
  \bibnamefont {Lam}}, \bibinfo {author} {\bibfnamefont {A.}~\bibnamefont
  {Coogan}}, \bibinfo {author} {\bibfnamefont {D.}~\bibnamefont
  {Foreman-Mackey}}, \bibinfo {author} {\bibfnamefont {M.}~\bibnamefont {Isi}},
  \ and\ \bibinfo {author} {\bibfnamefont {A.}~\bibnamefont {Zimmerman}},\
  }\href@noop {} {\  (\bibinfo {year} {2023})},\ \Eprint
  {http://arxiv.org/abs/2302.05329} {arXiv:2302.05329 [astro-ph.IM]}
  \BibitemShut {NoStop}%
\bibitem [{\citenamefont {Wong}\ \emph {et~al.}(2023)\citenamefont {Wong},
  \citenamefont {Isi},\ and\ \citenamefont {Edwards}}]{Wong:2023lgb}%
  \BibitemOpen
  \bibfield  {author} {\bibinfo {author} {\bibfnamefont {K.~W.~K.}\
  \bibnamefont {Wong}}, \bibinfo {author} {\bibfnamefont {M.}~\bibnamefont
  {Isi}}, \ and\ \bibinfo {author} {\bibfnamefont {T.~D.~P.}\ \bibnamefont
  {Edwards}},\ }\href@noop {} {\  (\bibinfo {year} {2023})},\ \Eprint
  {http://arxiv.org/abs/2302.05333} {arXiv:2302.05333 [astro-ph.IM]}
  \BibitemShut {NoStop}%
\bibitem [{\citenamefont {Zackay}\ \emph {et~al.}(2018)\citenamefont {Zackay},
  \citenamefont {Dai},\ and\ \citenamefont {Venumadhav}}]{Zackay:2018qdy}%
  \BibitemOpen
  \bibfield  {author} {\bibinfo {author} {\bibfnamefont {B.}~\bibnamefont
  {Zackay}}, \bibinfo {author} {\bibfnamefont {L.}~\bibnamefont {Dai}}, \ and\
  \bibinfo {author} {\bibfnamefont {T.}~\bibnamefont {Venumadhav}},\
  }\href@noop {} {\  (\bibinfo {year} {2018})},\ \Eprint
  {http://arxiv.org/abs/1806.08792} {arXiv:1806.08792 [astro-ph.IM]}
  \BibitemShut {NoStop}%
\bibitem [{\citenamefont {Leslie}\ \emph {et~al.}(2021)\citenamefont {Leslie},
  \citenamefont {Dai},\ and\ \citenamefont {Pratten}}]{Leslie:2021ssu}%
  \BibitemOpen
  \bibfield  {author} {\bibinfo {author} {\bibfnamefont {N.}~\bibnamefont
  {Leslie}}, \bibinfo {author} {\bibfnamefont {L.}~\bibnamefont {Dai}}, \ and\
  \bibinfo {author} {\bibfnamefont {G.}~\bibnamefont {Pratten}},\ }\href
  {\doibase 10.1103/PhysRevD.104.123030} {\bibfield  {journal} {\bibinfo
  {journal} {Phys. Rev. D}\ }\textbf {\bibinfo {volume} {104}},\ \bibinfo
  {pages} {123030} (\bibinfo {year} {2021})},\ \Eprint
  {http://arxiv.org/abs/2109.09872} {arXiv:2109.09872 [astro-ph.IM]}
  \BibitemShut {NoStop}%
\bibitem [{\citenamefont {Cranmer}\ \emph {et~al.}(2020)\citenamefont
  {Cranmer}, \citenamefont {Brehmer},\ and\ \citenamefont
  {Louppe}}]{Cranmer:2019eaq}%
  \BibitemOpen
  \bibfield  {author} {\bibinfo {author} {\bibfnamefont {K.}~\bibnamefont
  {Cranmer}}, \bibinfo {author} {\bibfnamefont {J.}~\bibnamefont {Brehmer}}, \
  and\ \bibinfo {author} {\bibfnamefont {G.}~\bibnamefont {Louppe}},\ }\href
  {\doibase 10.1073/pnas.1912789117} {\bibfield  {journal} {\bibinfo  {journal}
  {Proc. Nat. Acad. Sci.}\ }\textbf {\bibinfo {volume} {117}},\ \bibinfo
  {pages} {30055} (\bibinfo {year} {2020})},\ \Eprint
  {http://arxiv.org/abs/1911.01429} {arXiv:1911.01429 [stat.ML]} \BibitemShut
  {NoStop}%
\bibitem [{\citenamefont {Miller}\ \emph {et~al.}(2021)\citenamefont {Miller},
  \citenamefont {Cole}, \citenamefont {Forr\'e}, \citenamefont {Louppe},\ and\
  \citenamefont {Weniger}}]{Miller:2021hys}%
  \BibitemOpen
  \bibfield  {author} {\bibinfo {author} {\bibfnamefont {B.~K.}\ \bibnamefont
  {Miller}}, \bibinfo {author} {\bibfnamefont {A.}~\bibnamefont {Cole}},
  \bibinfo {author} {\bibfnamefont {P.}~\bibnamefont {Forr\'e}}, \bibinfo
  {author} {\bibfnamefont {G.}~\bibnamefont {Louppe}}, \ and\ \bibinfo {author}
  {\bibfnamefont {C.}~\bibnamefont {Weniger}},\ }in\ \href {\doibase
  10.5281/zenodo.5043706} {\emph {\bibinfo {booktitle} {{35th Conference on
  Neural Information Processing Systems}}}}\ (\bibinfo {year} {2021})\ \Eprint
  {http://arxiv.org/abs/2107.01214} {arXiv:2107.01214 [stat.ML]} \BibitemShut
  {NoStop}%
\bibitem [{\citenamefont {Tejero-Cantero}\ \emph {et~al.}(2020)\citenamefont
  {Tejero-Cantero}, \citenamefont {Boelts}, \citenamefont {Deistler},
  \citenamefont {Lueckmann}, \citenamefont {Durkan}, \citenamefont
  {Gonçalves}, \citenamefont {Greenberg},\ and\ \citenamefont
  {Macke}}]{Tejero-cantero:2020aaa}%
  \BibitemOpen
  \bibfield  {author} {\bibinfo {author} {\bibfnamefont {A.}~\bibnamefont
  {Tejero-Cantero}}, \bibinfo {author} {\bibfnamefont {J.}~\bibnamefont
  {Boelts}}, \bibinfo {author} {\bibfnamefont {M.}~\bibnamefont {Deistler}},
  \bibinfo {author} {\bibfnamefont {J.-M.}\ \bibnamefont {Lueckmann}}, \bibinfo
  {author} {\bibfnamefont {C.}~\bibnamefont {Durkan}}, \bibinfo {author}
  {\bibfnamefont {P.~J.}\ \bibnamefont {Gonçalves}}, \bibinfo {author}
  {\bibfnamefont {D.~S.}\ \bibnamefont {Greenberg}}, \ and\ \bibinfo {author}
  {\bibfnamefont {J.~H.}\ \bibnamefont {Macke}},\ }\href {\doibase
  10.21105/joss.02505} {\bibfield  {journal} {\bibinfo  {journal} {Journal of
  Open Source Software}\ }\textbf {\bibinfo {volume} {5}},\ \bibinfo {pages}
  {2505} (\bibinfo {year} {2020})}\BibitemShut {NoStop}%
\bibitem [{\citenamefont {Alsing}\ \emph {et~al.}(2019)\citenamefont {Alsing},
  \citenamefont {Charnock}, \citenamefont {Feeney},\ and\ \citenamefont
  {Wandelt}}]{Alsing:2019xrx}%
  \BibitemOpen
  \bibfield  {author} {\bibinfo {author} {\bibfnamefont {J.}~\bibnamefont
  {Alsing}}, \bibinfo {author} {\bibfnamefont {T.}~\bibnamefont {Charnock}},
  \bibinfo {author} {\bibfnamefont {S.}~\bibnamefont {Feeney}}, \ and\ \bibinfo
  {author} {\bibfnamefont {B.}~\bibnamefont {Wandelt}},\ }\href {\doibase
  10.1093/mnras/stz1960} {\bibfield  {journal} {\bibinfo  {journal} {Mon. Not.
  Roy. Astron. Soc.}\ }\textbf {\bibinfo {volume} {488}},\ \bibinfo {pages}
  {4440} (\bibinfo {year} {2019})},\ \Eprint {http://arxiv.org/abs/1903.00007}
  {arXiv:1903.00007 [astro-ph.CO]} \BibitemShut {NoStop}%
\bibitem [{\citenamefont {Cole}\ \emph {et~al.}(2022)\citenamefont {Cole},
  \citenamefont {Miller}, \citenamefont {Witte}, \citenamefont {Cai},
  \citenamefont {Grootes}, \citenamefont {Nattino},\ and\ \citenamefont
  {Weniger}}]{Cole:2021gwr}%
  \BibitemOpen
  \bibfield  {author} {\bibinfo {author} {\bibfnamefont {A.}~\bibnamefont
  {Cole}}, \bibinfo {author} {\bibfnamefont {B.~K.}\ \bibnamefont {Miller}},
  \bibinfo {author} {\bibfnamefont {S.~J.}\ \bibnamefont {Witte}}, \bibinfo
  {author} {\bibfnamefont {M.~X.}\ \bibnamefont {Cai}}, \bibinfo {author}
  {\bibfnamefont {M.~W.}\ \bibnamefont {Grootes}}, \bibinfo {author}
  {\bibfnamefont {F.}~\bibnamefont {Nattino}}, \ and\ \bibinfo {author}
  {\bibfnamefont {C.}~\bibnamefont {Weniger}},\ }\href {\doibase
  10.1088/1475-7516/2022/09/004} {\bibfield  {journal} {\bibinfo  {journal}
  {JCAP}\ }\textbf {\bibinfo {volume} {09}},\ \bibinfo {pages} {004} (\bibinfo
  {year} {2022})},\ \Eprint {http://arxiv.org/abs/2111.08030} {arXiv:2111.08030
  [astro-ph.CO]} \BibitemShut {NoStop}%
\bibitem [{\citenamefont {Montel}\ \emph {et~al.}(2022)\citenamefont {Montel},
  \citenamefont {Coogan}, \citenamefont {Correa}, \citenamefont {Karchev},\
  and\ \citenamefont {Weniger}}]{Montel:2022fhv}%
  \BibitemOpen
  \bibfield  {author} {\bibinfo {author} {\bibfnamefont {N.~A.}\ \bibnamefont
  {Montel}}, \bibinfo {author} {\bibfnamefont {A.}~\bibnamefont {Coogan}},
  \bibinfo {author} {\bibfnamefont {C.}~\bibnamefont {Correa}}, \bibinfo
  {author} {\bibfnamefont {K.}~\bibnamefont {Karchev}}, \ and\ \bibinfo
  {author} {\bibfnamefont {C.}~\bibnamefont {Weniger}},\ }\href {\doibase
  10.1093/mnras/stac3215} {\bibfield  {journal} {\bibinfo  {journal} {Mon. Not.
  Roy. Astron. Soc.}\ }\textbf {\bibinfo {volume} {518}},\ \bibinfo {pages}
  {2746} (\bibinfo {year} {2022})},\ \Eprint {http://arxiv.org/abs/2205.09126}
  {arXiv:2205.09126 [astro-ph.CO]} \BibitemShut {NoStop}%
\bibitem [{\citenamefont {Anau~Montel}\ and\ \citenamefont
  {Weniger}(2022)}]{AnauMontel:2022ppb}%
  \BibitemOpen
  \bibfield  {author} {\bibinfo {author} {\bibfnamefont {N.}~\bibnamefont
  {Anau~Montel}}\ and\ \bibinfo {author} {\bibfnamefont {C.}~\bibnamefont
  {Weniger}},\ }in\ \href@noop {} {\emph {\bibinfo {booktitle} {{36th
  Conference on Neural Information Processing Systems}}}}\ (\bibinfo {year}
  {2022})\ \Eprint {http://arxiv.org/abs/2211.04291} {arXiv:2211.04291
  [astro-ph.IM]} \BibitemShut {NoStop}%
\bibitem [{\citenamefont {Dimitriou}\ \emph {et~al.}(2022)\citenamefont
  {Dimitriou}, \citenamefont {Weniger},\ and\ \citenamefont
  {Correa}}]{Dimitriou:2022cvc}%
  \BibitemOpen
  \bibfield  {author} {\bibinfo {author} {\bibfnamefont {A.}~\bibnamefont
  {Dimitriou}}, \bibinfo {author} {\bibfnamefont {C.}~\bibnamefont {Weniger}},
  \ and\ \bibinfo {author} {\bibfnamefont {C.~A.}\ \bibnamefont {Correa}},\
  }\href@noop {} {\  (\bibinfo {year} {2022})},\ \Eprint
  {http://arxiv.org/abs/2206.11312} {arXiv:2206.11312 [astro-ph.CO]}
  \BibitemShut {NoStop}%
\bibitem [{\citenamefont {Karchev}\ \emph {et~al.}(2022)\citenamefont
  {Karchev}, \citenamefont {Trotta},\ and\ \citenamefont
  {Weniger}}]{Karchev:2022xyn}%
  \BibitemOpen
  \bibfield  {author} {\bibinfo {author} {\bibfnamefont {K.}~\bibnamefont
  {Karchev}}, \bibinfo {author} {\bibfnamefont {R.}~\bibnamefont {Trotta}}, \
  and\ \bibinfo {author} {\bibfnamefont {C.}~\bibnamefont {Weniger}},\ }\href
  {\doibase 10.1093/mnras/stac3785} {\bibfield  {journal} {\bibinfo  {journal}
  {{Mon. Not. Roy. Astron. Soc.}}\ }\textbf {\bibinfo {volume} {520}},\
  \bibinfo {pages} {1056} (\bibinfo {year} {2022})},\ \Eprint
  {http://arxiv.org/abs/2209.06733} {arXiv:2209.06733 [astro-ph.CO]}
  \BibitemShut {NoStop}%
\bibitem [{\citenamefont {Alvey}\ \emph {et~al.}(2023)\citenamefont {Alvey},
  \citenamefont {Gerdes},\ and\ \citenamefont {Weniger}}]{Alvey:2023pkx}%
  \BibitemOpen
  \bibfield  {author} {\bibinfo {author} {\bibfnamefont {J.}~\bibnamefont
  {Alvey}}, \bibinfo {author} {\bibfnamefont {M.}~\bibnamefont {Gerdes}}, \
  and\ \bibinfo {author} {\bibfnamefont {C.}~\bibnamefont {Weniger}},\
  }\href@noop {} {\  (\bibinfo {year} {2023})},\ \Eprint
  {http://arxiv.org/abs/2304.02032} {arXiv:2304.02032 [astro-ph.GA]}
  \BibitemShut {NoStop}%
\bibitem [{\citenamefont {Dax}\ \emph {et~al.}(2021{\natexlab{a}})\citenamefont
  {Dax}, \citenamefont {Green}, \citenamefont {Gair}, \citenamefont {Deistler},
  \citenamefont {Sch\"olkopf},\ and\ \citenamefont {Macke}}]{Dax:2021myb}%
  \BibitemOpen
  \bibfield  {author} {\bibinfo {author} {\bibfnamefont {M.}~\bibnamefont
  {Dax}}, \bibinfo {author} {\bibfnamefont {S.~R.}\ \bibnamefont {Green}},
  \bibinfo {author} {\bibfnamefont {J.}~\bibnamefont {Gair}}, \bibinfo {author}
  {\bibfnamefont {M.}~\bibnamefont {Deistler}}, \bibinfo {author}
  {\bibfnamefont {B.}~\bibnamefont {Sch\"olkopf}}, \ and\ \bibinfo {author}
  {\bibfnamefont {J.~H.}\ \bibnamefont {Macke}},\ }\href@noop {} {\  (\bibinfo
  {year} {2021}{\natexlab{a}})},\ \Eprint {http://arxiv.org/abs/2111.13139}
  {arXiv:2111.13139 [cs.LG]} \BibitemShut {NoStop}%
\bibitem [{\citenamefont {Dax}\ \emph {et~al.}(2021{\natexlab{b}})\citenamefont
  {Dax}, \citenamefont {Green}, \citenamefont {Gair}, \citenamefont {Macke},
  \citenamefont {Buonanno},\ and\ \citenamefont {Sch\"olkopf}}]{Dax:2021tsq}%
  \BibitemOpen
  \bibfield  {author} {\bibinfo {author} {\bibfnamefont {M.}~\bibnamefont
  {Dax}}, \bibinfo {author} {\bibfnamefont {S.~R.}\ \bibnamefont {Green}},
  \bibinfo {author} {\bibfnamefont {J.}~\bibnamefont {Gair}}, \bibinfo {author}
  {\bibfnamefont {J.~H.}\ \bibnamefont {Macke}}, \bibinfo {author}
  {\bibfnamefont {A.}~\bibnamefont {Buonanno}}, \ and\ \bibinfo {author}
  {\bibfnamefont {B.}~\bibnamefont {Sch\"olkopf}},\ }\href {\doibase
  10.1103/PhysRevLett.127.241103} {\bibfield  {journal} {\bibinfo  {journal}
  {Phys. Rev. Lett.}\ }\textbf {\bibinfo {volume} {127}},\ \bibinfo {pages}
  {241103} (\bibinfo {year} {2021}{\natexlab{b}})},\ \Eprint
  {http://arxiv.org/abs/2106.12594} {arXiv:2106.12594 [gr-qc]} \BibitemShut
  {NoStop}%
\bibitem [{\citenamefont {Dax}\ \emph {et~al.}(2023)\citenamefont {Dax},
  \citenamefont {Green}, \citenamefont {Gair}, \citenamefont {P\"urrer},
  \citenamefont {Wildberger}, \citenamefont {Macke}, \citenamefont {Buonanno},\
  and\ \citenamefont {Sch\"olkopf}}]{Dax:2022pxd}%
  \BibitemOpen
  \bibfield  {author} {\bibinfo {author} {\bibfnamefont {M.}~\bibnamefont
  {Dax}}, \bibinfo {author} {\bibfnamefont {S.~R.}\ \bibnamefont {Green}},
  \bibinfo {author} {\bibfnamefont {J.}~\bibnamefont {Gair}}, \bibinfo {author}
  {\bibfnamefont {M.}~\bibnamefont {P\"urrer}}, \bibinfo {author}
  {\bibfnamefont {J.}~\bibnamefont {Wildberger}}, \bibinfo {author}
  {\bibfnamefont {J.~H.}\ \bibnamefont {Macke}}, \bibinfo {author}
  {\bibfnamefont {A.}~\bibnamefont {Buonanno}}, \ and\ \bibinfo {author}
  {\bibfnamefont {B.}~\bibnamefont {Sch\"olkopf}},\ }\href {\doibase
  10.1103/PhysRevLett.130.171403} {\bibfield  {journal} {\bibinfo  {journal}
  {Phys. Rev. Lett.}\ }\textbf {\bibinfo {volume} {130}},\ \bibinfo {pages}
  {171403} (\bibinfo {year} {2023})},\ \Eprint
  {http://arxiv.org/abs/2210.05686} {arXiv:2210.05686 [gr-qc]} \BibitemShut
  {NoStop}%
\bibitem [{\citenamefont {Miller}\ \emph {et~al.}(2022)\citenamefont {Miller},
  \citenamefont {Cole}, \citenamefont {Weniger}, \citenamefont {Nattino},
  \citenamefont {Ku},\ and\ \citenamefont {Grootes}}]{Miller:2022shs}%
  \BibitemOpen
  \bibfield  {author} {\bibinfo {author} {\bibfnamefont {B.~K.}\ \bibnamefont
  {Miller}}, \bibinfo {author} {\bibfnamefont {A.}~\bibnamefont {Cole}},
  \bibinfo {author} {\bibfnamefont {C.}~\bibnamefont {Weniger}}, \bibinfo
  {author} {\bibfnamefont {F.}~\bibnamefont {Nattino}}, \bibinfo {author}
  {\bibfnamefont {O.}~\bibnamefont {Ku}}, \ and\ \bibinfo {author}
  {\bibfnamefont {M.~W.}\ \bibnamefont {Grootes}},\ }\href {\doibase
  10.21105/joss.04205} {\bibfield  {journal} {\bibinfo  {journal} {J. Open
  Source Softw.}\ }\textbf {\bibinfo {volume} {7}},\ \bibinfo {pages} {4205}
  (\bibinfo {year} {2022})}\BibitemShut {NoStop}%
\bibitem [{\citenamefont {Bhardwaj}\ \emph {et~al.}(2023)\citenamefont
  {Bhardwaj}, \citenamefont {Alvey}, \citenamefont {Miller}, \citenamefont
  {Nissanke},\ and\ \citenamefont {Weniger}}]{Bhardwaj:2023xph}%
  \BibitemOpen
  \bibfield  {author} {\bibinfo {author} {\bibfnamefont {U.}~\bibnamefont
  {Bhardwaj}}, \bibinfo {author} {\bibfnamefont {J.}~\bibnamefont {Alvey}},
  \bibinfo {author} {\bibfnamefont {B.~K.}\ \bibnamefont {Miller}}, \bibinfo
  {author} {\bibfnamefont {S.}~\bibnamefont {Nissanke}}, \ and\ \bibinfo
  {author} {\bibfnamefont {C.}~\bibnamefont {Weniger}},\ }\href@noop {} {\
  (\bibinfo {year} {2023})},\ \Eprint {http://arxiv.org/abs/2304.02035}
  {arXiv:2304.02035 [gr-qc]} \BibitemShut {NoStop}%
\bibitem [{\citenamefont {Ashton}\ and\ \citenamefont
  {Talbot}(2021)}]{Ashton:2021anp}%
  \BibitemOpen
  \bibfield  {author} {\bibinfo {author} {\bibfnamefont {G.}~\bibnamefont
  {Ashton}}\ and\ \bibinfo {author} {\bibfnamefont {C.}~\bibnamefont
  {Talbot}},\ }\href {\doibase 10.1093/mnras/stab2236} {\bibfield  {journal}
  {\bibinfo  {journal} {Mon. Not. Roy. Astron. Soc.}\ }\textbf {\bibinfo
  {volume} {507}},\ \bibinfo {pages} {2037} (\bibinfo {year} {2021})},\ \Eprint
  {http://arxiv.org/abs/2106.08730} {arXiv:2106.08730 [gr-qc]} \BibitemShut
  {NoStop}%
\bibitem [{\citenamefont {Abbott}\ \emph {et~al.}(2020)\citenamefont {Abbott}
  \emph {et~al.}}]{LIGOScientific:2019hgc}%
  \BibitemOpen
  \bibfield  {author} {\bibinfo {author} {\bibfnamefont {B.~P.}\ \bibnamefont
  {Abbott}} \emph {et~al.} (\bibinfo {collaboration} {LIGO Scientific,
  Virgo}),\ }\href {\doibase 10.1088/1361-6382/ab685e} {\bibfield  {journal}
  {\bibinfo  {journal} {Class. Quant. Grav.}\ }\textbf {\bibinfo {volume}
  {37}},\ \bibinfo {pages} {055002} (\bibinfo {year} {2020})},\ \Eprint
  {http://arxiv.org/abs/1908.11170} {arXiv:1908.11170 [gr-qc]} \BibitemShut
  {NoStop}%
\bibitem [{\citenamefont {Pratten}\ \emph {et~al.}(2021)\citenamefont {Pratten}
  \emph {et~al.}}]{Pratten:2020ceb}%
  \BibitemOpen
  \bibfield  {author} {\bibinfo {author} {\bibfnamefont {G.}~\bibnamefont
  {Pratten}} \emph {et~al.},\ }\href {\doibase 10.1103/PhysRevD.103.104056}
  {\bibfield  {journal} {\bibinfo  {journal} {Phys. Rev. D}\ }\textbf {\bibinfo
  {volume} {103}},\ \bibinfo {pages} {104056} (\bibinfo {year} {2021})},\
  \Eprint {http://arxiv.org/abs/2004.06503} {arXiv:2004.06503 [gr-qc]}
  \BibitemShut {NoStop}%
\bibitem [{\citenamefont {Green}(2021)}]{Green:2021stw}%
  \BibitemOpen
  \bibfield  {author} {\bibinfo {author} {\bibfnamefont {R.~D.}\ \bibnamefont
  {Green}},\ }\emph {\bibinfo {title} {{There\textquoteright{}s more than one
  way to ride the wave: A multi-disciplinary approach to gravitational wave
  data analysis}}},\ \href@noop {} {Ph.D. thesis},\ \bibinfo  {school} {Cardiff
  U.} (\bibinfo {year} {2021})\BibitemShut {NoStop}%
\bibitem [{\citenamefont {Kim}\ and\ \citenamefont {Liu}(2023)}]{Kim:2023scq}%
  \BibitemOpen
  \bibfield  {author} {\bibinfo {author} {\bibfnamefont {K.}~\bibnamefont
  {Kim}}\ and\ \bibinfo {author} {\bibfnamefont {A.}~\bibnamefont {Liu}},\
  }\href@noop {} {\  (\bibinfo {year} {2023})},\ \Eprint
  {http://arxiv.org/abs/2301.07253} {arXiv:2301.07253 [gr-qc]} \BibitemShut
  {NoStop}%
\bibitem [{\citenamefont {Taracchini}\ \emph {et~al.}(2014)\citenamefont
  {Taracchini} \emph {et~al.}}]{Taracchini:2013rva}%
  \BibitemOpen
  \bibfield  {author} {\bibinfo {author} {\bibfnamefont {A.}~\bibnamefont
  {Taracchini}} \emph {et~al.},\ }\href {\doibase 10.1103/PhysRevD.89.061502}
  {\bibfield  {journal} {\bibinfo  {journal} {Phys. Rev. D}\ }\textbf {\bibinfo
  {volume} {89}},\ \bibinfo {pages} {061502} (\bibinfo {year} {2014})},\
  \Eprint {http://arxiv.org/abs/1311.2544} {arXiv:1311.2544 [gr-qc]}
  \BibitemShut {NoStop}%
\bibitem [{\citenamefont {Boh\'e}\ \emph {et~al.}(2017)\citenamefont {Boh\'e}
  \emph {et~al.}}]{Bohe:2016gbl}%
  \BibitemOpen
  \bibfield  {author} {\bibinfo {author} {\bibfnamefont {A.}~\bibnamefont
  {Boh\'e}} \emph {et~al.},\ }\href {\doibase 10.1103/PhysRevD.95.044028}
  {\bibfield  {journal} {\bibinfo  {journal} {Phys. Rev. D}\ }\textbf {\bibinfo
  {volume} {95}},\ \bibinfo {pages} {044028} (\bibinfo {year} {2017})},\
  \Eprint {http://arxiv.org/abs/1611.03703} {arXiv:1611.03703 [gr-qc]}
  \BibitemShut {NoStop}%
\bibitem [{\citenamefont {Ossokine}\ \emph {et~al.}(2020)\citenamefont
  {Ossokine} \emph {et~al.}}]{Ossokine:2020kjp}%
  \BibitemOpen
  \bibfield  {author} {\bibinfo {author} {\bibfnamefont {S.}~\bibnamefont
  {Ossokine}} \emph {et~al.},\ }\href {\doibase 10.1103/PhysRevD.102.044055}
  {\bibfield  {journal} {\bibinfo  {journal} {Phys. Rev. D}\ }\textbf {\bibinfo
  {volume} {102}},\ \bibinfo {pages} {044055} (\bibinfo {year} {2020})},\
  \Eprint {http://arxiv.org/abs/2004.09442} {arXiv:2004.09442 [gr-qc]}
  \BibitemShut {NoStop}%
\bibitem [{\citenamefont {Khan}\ \emph {et~al.}(2016)\citenamefont {Khan},
  \citenamefont {Husa}, \citenamefont {Hannam}, \citenamefont {Ohme},
  \citenamefont {P\"urrer}, \citenamefont {Jim\'enez~Forteza},\ and\
  \citenamefont {Boh\'e}}]{Khan:2015jqa}%
  \BibitemOpen
  \bibfield  {author} {\bibinfo {author} {\bibfnamefont {S.}~\bibnamefont
  {Khan}}, \bibinfo {author} {\bibfnamefont {S.}~\bibnamefont {Husa}}, \bibinfo
  {author} {\bibfnamefont {M.}~\bibnamefont {Hannam}}, \bibinfo {author}
  {\bibfnamefont {F.}~\bibnamefont {Ohme}}, \bibinfo {author} {\bibfnamefont
  {M.}~\bibnamefont {P\"urrer}}, \bibinfo {author} {\bibfnamefont
  {X.}~\bibnamefont {Jim\'enez~Forteza}}, \ and\ \bibinfo {author}
  {\bibfnamefont {A.}~\bibnamefont {Boh\'e}},\ }\href {\doibase
  10.1103/PhysRevD.93.044007} {\bibfield  {journal} {\bibinfo  {journal} {Phys.
  Rev. D}\ }\textbf {\bibinfo {volume} {93}},\ \bibinfo {pages} {044007}
  (\bibinfo {year} {2016})},\ \Eprint {http://arxiv.org/abs/1508.07253}
  {arXiv:1508.07253 [gr-qc]} \BibitemShut {NoStop}%
\bibitem [{\citenamefont {Hannam}\ \emph {et~al.}(2014)\citenamefont {Hannam},
  \citenamefont {Schmidt}, \citenamefont {Boh\'e}, \citenamefont {Haegel},
  \citenamefont {Husa}, \citenamefont {Ohme}, \citenamefont {Pratten},\ and\
  \citenamefont {P\"urrer}}]{Hannam:2013oca}%
  \BibitemOpen
  \bibfield  {author} {\bibinfo {author} {\bibfnamefont {M.}~\bibnamefont
  {Hannam}}, \bibinfo {author} {\bibfnamefont {P.}~\bibnamefont {Schmidt}},
  \bibinfo {author} {\bibfnamefont {A.}~\bibnamefont {Boh\'e}}, \bibinfo
  {author} {\bibfnamefont {L.}~\bibnamefont {Haegel}}, \bibinfo {author}
  {\bibfnamefont {S.}~\bibnamefont {Husa}}, \bibinfo {author} {\bibfnamefont
  {F.}~\bibnamefont {Ohme}}, \bibinfo {author} {\bibfnamefont {G.}~\bibnamefont
  {Pratten}}, \ and\ \bibinfo {author} {\bibfnamefont {M.}~\bibnamefont
  {P\"urrer}},\ }\href {\doibase 10.1103/PhysRevLett.113.151101} {\bibfield
  {journal} {\bibinfo  {journal} {Phys. Rev. Lett.}\ }\textbf {\bibinfo
  {volume} {113}},\ \bibinfo {pages} {151101} (\bibinfo {year} {2014})},\
  \Eprint {http://arxiv.org/abs/1308.3271} {arXiv:1308.3271 [gr-qc]}
  \BibitemShut {NoStop}%
\bibitem [{\citenamefont {{Hermans}}\ \emph {et~al.}(2021)\citenamefont
  {{Hermans}}, \citenamefont {{Delaunoy}}, \citenamefont {{Rozet}},
  \citenamefont {{Wehenkel}}, \citenamefont {{Begy}},\ and\ \citenamefont
  {{Louppe}}}]{Hermans:2021aaa}%
  \BibitemOpen
  \bibfield  {author} {\bibinfo {author} {\bibfnamefont {J.}~\bibnamefont
  {{Hermans}}}, \bibinfo {author} {\bibfnamefont {A.}~\bibnamefont
  {{Delaunoy}}}, \bibinfo {author} {\bibfnamefont {F.}~\bibnamefont {{Rozet}}},
  \bibinfo {author} {\bibfnamefont {A.}~\bibnamefont {{Wehenkel}}}, \bibinfo
  {author} {\bibfnamefont {V.}~\bibnamefont {{Begy}}}, \ and\ \bibinfo {author}
  {\bibfnamefont {G.}~\bibnamefont {{Louppe}}},\ }\href@noop {} {\  (\bibinfo
  {year} {2021})},\ \Eprint {http://arxiv.org/abs/2110.06581} {arXiv:2110.06581
  [stat.ML]} \BibitemShut {NoStop}%
\bibitem [{\citenamefont {{Lemos}}\ \emph {et~al.}(2023)\citenamefont
  {{Lemos}}, \citenamefont {{Coogan}}, \citenamefont {{Hezaveh}},\ and\
  \citenamefont {{Perreault-Levasseur}}}]{Lemos:2023aaa}%
  \BibitemOpen
  \bibfield  {author} {\bibinfo {author} {\bibfnamefont {P.}~\bibnamefont
  {{Lemos}}}, \bibinfo {author} {\bibfnamefont {A.}~\bibnamefont {{Coogan}}},
  \bibinfo {author} {\bibfnamefont {Y.}~\bibnamefont {{Hezaveh}}}, \ and\
  \bibinfo {author} {\bibfnamefont {L.}~\bibnamefont {{Perreault-Levasseur}}},\
  }\href@noop {} {\  (\bibinfo {year} {2023})},\ \Eprint
  {http://arxiv.org/abs/2302.03026} {arXiv:2302.03026 [stat.ML]} \BibitemShut
  {NoStop}%
\end{thebibliography}%
\clearpage
\newpage

\title{What to do when things get crowded? \\ Scalable joint analysis of overlapping gravitational wave signals \\ \textit{Supplemental Material}}
\maketitle

\onecolumngrid 
\twocolumngrid

\setcounter{figure}{0}
\renewcommand{\thefigure}{S\arabic{figure}}
\setcounter{table}{0}
\renewcommand{\thetable}{S\arabic{table}}
\vspace*{-40pt}
\section{Coverage Tests}

\vspace*{-10pt}
\noindent Coverage tests explicitly check the behaviour of the posterior over various realizations of the signal. The intuition is that if our posterior estimates are behaving correctly, then \textit{e.g.} for $p$\% of realizations, the true value of the parameter(s) should lie inside the $p$\% confidence interval. If our posteriors are too overconfident, then this will be true for more that $p$\% of the realizations, and vice versa for under-confident inference results. For each case study, to carry out these tests we generated an additional 2000 simulations from the truncated prior in the final round. Since our sequential SBI framework achieves local amortization, inference on these $2000$ signals needed to produce the coverage plots is almost immediate, requiring only a single network evaluation for each example. The full set of these tests are shown in Figs.~\ref{fig:C1_pp_full}--\ref{fig:C3_pp_full}.

\vspace*{-16pt}
\section{Relevant details of the \texttt{peregrine} implementation}\label{sec:network}

\vspace*{-10pt}
\noindent As discussed in the main text, the majority of the \texttt{peregrine} implementation described in Ref.~\cite{Bhardwaj:2023xph} has remained unchanged for this application to overlapping signals. In this supplementary section, we remind the reader of the key components and indicate any places where slight changes were made compared to this reference. 

\begin{figure*}[t]
    \centering
    \includegraphics[width=0.8\textwidth,trim={0.1cm 0.1cm 0.1cm 0.1cm},clip]{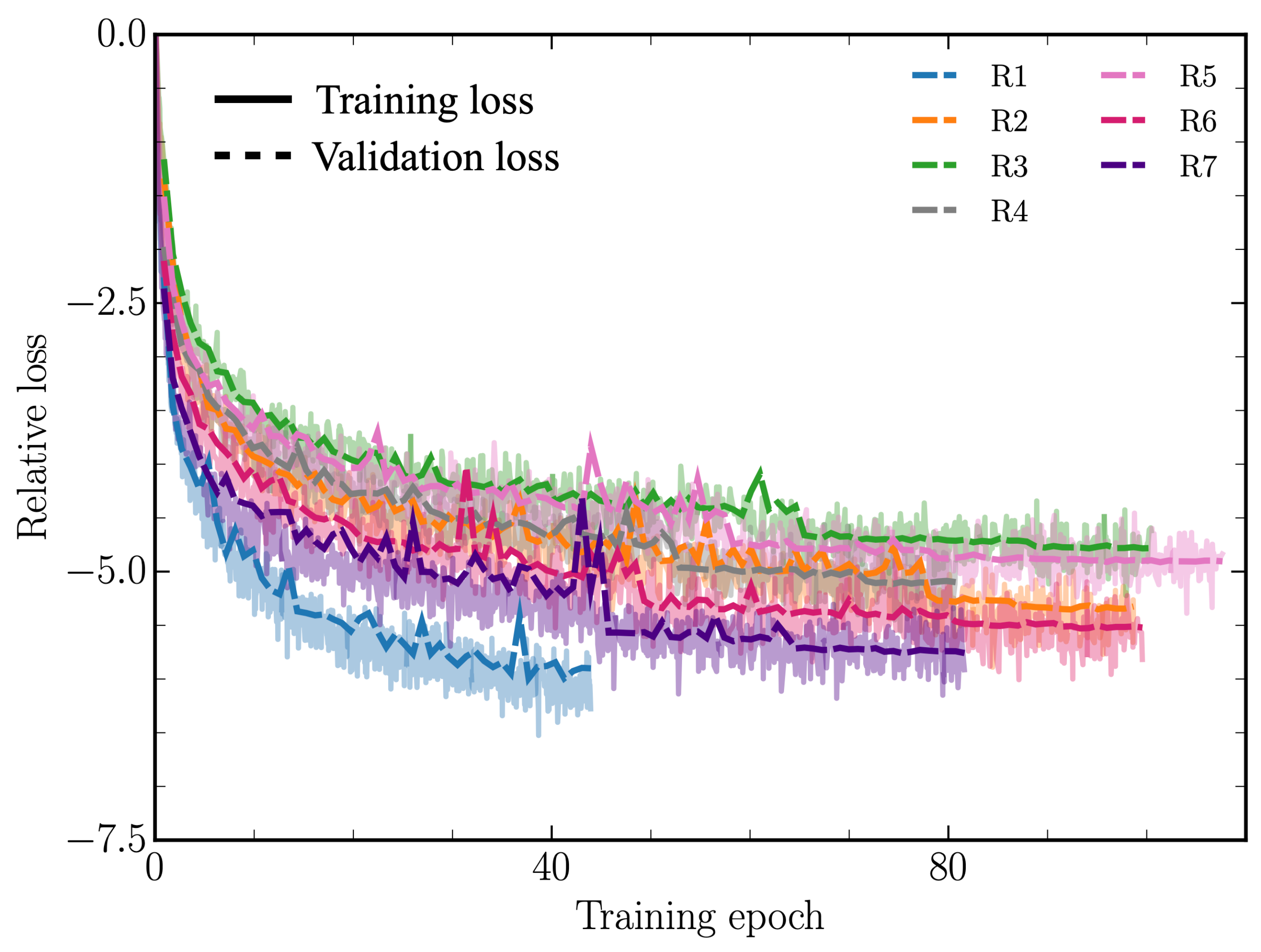}
    \caption{Training and validation loss curves for the various rounds of inference performed in this work. Values of the loss on the training dataset are shown by the solid lines, whilst the values on the validation set are shown by dashed curves. The example given here is for the first case study (\textbf{C1}) described in the main text, but a similar picture holds across all cases.}
    \label{fig:loss_curves}
\end{figure*}

The \texttt{peregrine} library is built on top of the \texttt{swyft}~\cite{Miller:2021hys, Miller:2022shs} implementation of the truncated marginal neural ratio estimation (TMNRE) algorithm. For this work, we use exactly the same \texttt{pytorch}-based embedding network described in the Appendix and Sec.~II.3 of Ref.~\cite{Bhardwaj:2023xph}. In addition, we use the same truncation scheme detailed in Sec.~II.2 of the same reference, with $\epsilon = 10^{-5}$, as defined in Refs.~\cite{Bhardwaj:2023xph,Miller:2021hys}. Furthermore, we use almost identical settings for the TMNRE algorithm provided in Tab.~II of~\cite{Bhardwaj:2023xph}. The only differences in this work are the simulation schedule and the training/validation batch size. In this work, we use $10^6$ simulations per round to account for the more complex parameter space (although we expect that this is overly conservative). We also take a correspondingly larger training/validation batch size of $512$ instead of $256$. As in the single signal analysis, we make use of the noise shuffling scheme explained in Sec.~II.3 of~\cite{Bhardwaj:2023xph}, which efficiently avoids overfitting. For the application in this work, we show the training and validation curves across the 7 rounds of inference in Fig.~\ref{fig:loss_curves}. We see that across all curves, the training proceeds smoothly and the training and validation curves are well calibrated. 

In terms of simulations, we generate waveform simulations using the waveform templates provided in \texttt{bilby}~\cite{Ashton:2018jfp}. In this paper, we use the \texttt{IMRPhenomXPHM}~\cite{Pratten:2020ceb} waveform approximant to generate both signals, however, any waveform choice is possible within our framework. For example, in Ref.~\cite{Bhardwaj:2023xph}, we instead used the \texttt{SEOBNRv4PHM} model~\cite{Ossokine:2020kjp}. For the case studies described in the main text, we generated signals of total duration $4$ s, which we explicitly checked was long enough (in the sense of the \texttt{bilby} functionality to compute \texttt{get\_safe\_signal\_duration}) for all parts of the prior parameter space.

\begin{figure}[t]
    \vspace{-10pt}
    \centering
    \includegraphics[width=0.95\columnwidth,trim={0.4cm 0.4cm 0.4cm 0.4cm},clip]{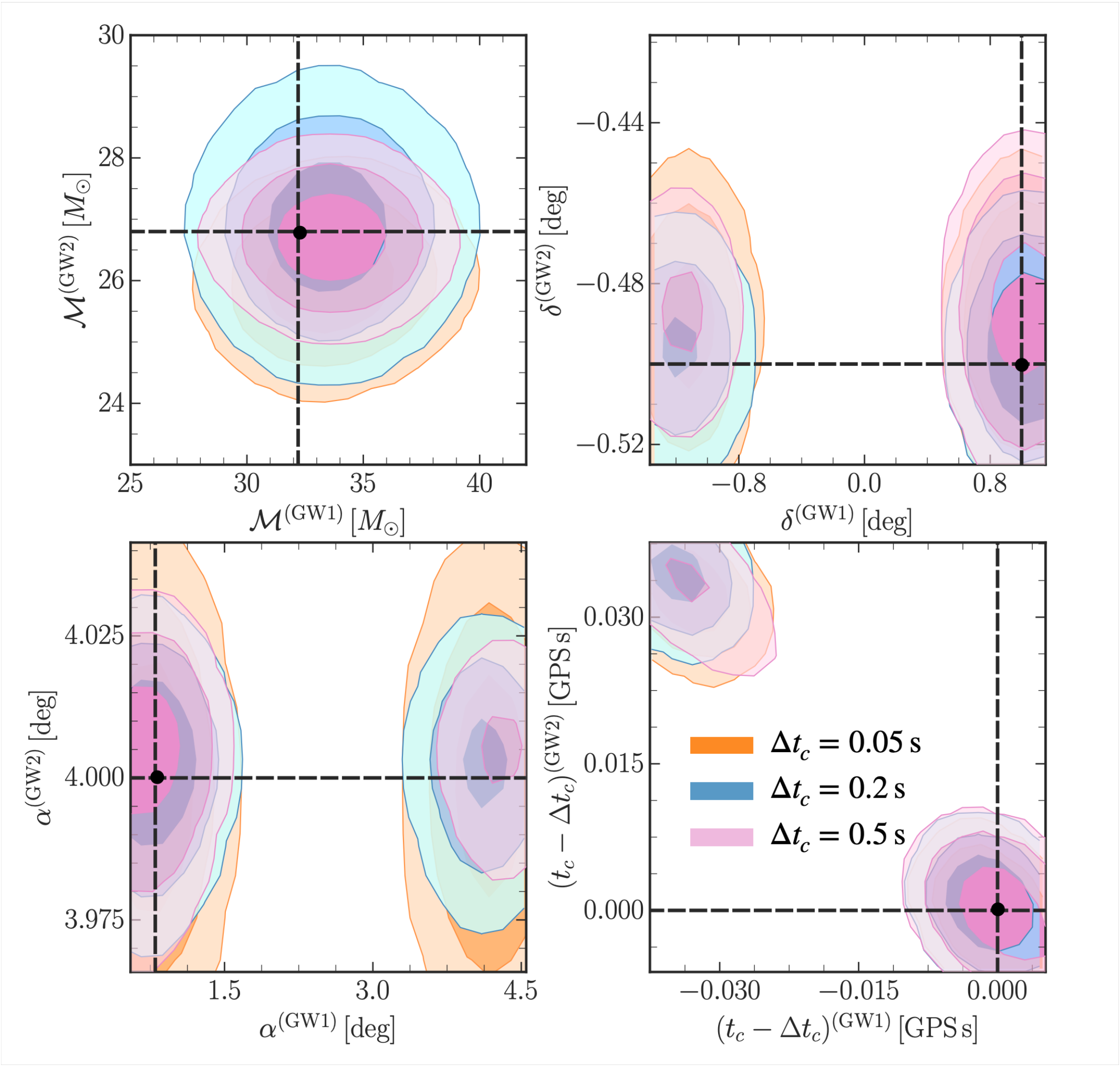}
    \caption{Additional 2D marginals for all three case studies (\textbf{C1}, \textbf{C2}, and \textbf{C3}). From top-left to bottom-right, we show the $\mathrm{2D}$ posteriors for the two chirp masses, declinations, right ascensions and merger times. Broadly, we see two features: there are few correlations between parameters from different signals, and that generally we achieve slightly better sensitivity as the merger times become more separated.}\vspace*{-16pt}
    \label{fig:2D_full}
\end{figure}

\begin{table*}[th!]
\centering
\resizebox{0.85\linewidth}{!}{
\begin{tabular}{@{}lccr@{}}
\hline
\textbf{Parameter}                                      & \textbf{Prior Choice}         & \textbf{Injection} (\textbf{GW1})   & \textbf{Injection} (\textbf{GW2})\\ \hline
Chirp mass $\mathcal{M}\,\mathrm{[M}_\odot\mathrm{]}$           & $\mathrm{U}(20, 100)$         & 32.2  & 26.8 \\
Mass ratio, $q$                                         & $\mathrm{U}(0.125, 1)$        & 0.77  & 0.81 \\
Inclination angle $\theta_{jn}\,\mathrm{[rad]}$         & $\mathrm{sine}(0, \pi)$ & 0.05  & 1.8 \\
Polarisation angle $\psi\,\mathrm{[rad]}$               & $\mathrm{U}(0, \pi)$          & 0.9  & 0.4 \\
Phase $\phi_c\,\mathrm{[rad]}$                          & $\mathrm{U}(0, 2\pi)$         & 1.5  & 0.6 \\
Tilt angles $\theta_{1}, \theta_2 \, \mathrm{[rad]}$    & $\mathrm{sine}(0, \pi)$ & 1.5, 1.5  & 0.9, 1.2 \\
Dimensionless spins $a_1, a_2$                          & $\mathrm{U}(0, 1)$         & 0.1, 0.05  & 0.7, 0.3 \\
Spin angles $\phi_{12}, \phi_{jl}\,\mathrm{[rad]}$      & $\mathrm{U}(0, 2\pi)$         & 0.8, 0.8  & 0.6, 0.9 \\
Right ascension $\alpha\,\mathrm{[rad]}$                & $\mathrm{U}(0, 2\pi)$         & 0.8  & 4.0 \\
Declination $\delta\,\mathrm{[rad]}$                    & $\mathrm{cosine}(-\pi/2, \pi/2)$ & 1.0  & -0.5 \\
Merger time $t_c\,\mathrm{[GPS\ s]}$                    & $\mathrm{U}(-0.1, 0.1)$       & 0.0  &  \\
Merger time difference $\Delta t_c\,\mathrm{[GPS\ s]}$                    & $\mathrm{U}(0, 1)$       &  &  0.05(\textbf{C1}), 0.2(\textbf{C2}), 0.5(\textbf{C3}) \\
Luminosity Distance $d_\mathrm{L}\,\mathrm{[Mpc]}$      & $\mathrm{U}_{\mathrm{vol.}}(100, 2500)^\star$ & 1545  & 636.5\\ \hline
\end{tabular}}
\caption{Definitions and prior choices for all relevant gravitational wave parameters in this work.\\
{\footnotesize ${}^\dagger$ Note that these are the default priors used in $\mathrm{bilby}$ analyses for BBH systems (subject to calibration ranges of the waveform approximants). \\
${}^\star$ Specifically, the luminosity distance prior is taken to be uniform in comoving volume in the \emph{source frame}.}}
\label{tab:sbi_params}
\end{table*}

\begin{figure*}
    \centering
    \includegraphics[width=\textwidth,trim={0.1cm 0.1cm 0.1cm 0.1cm},clip]{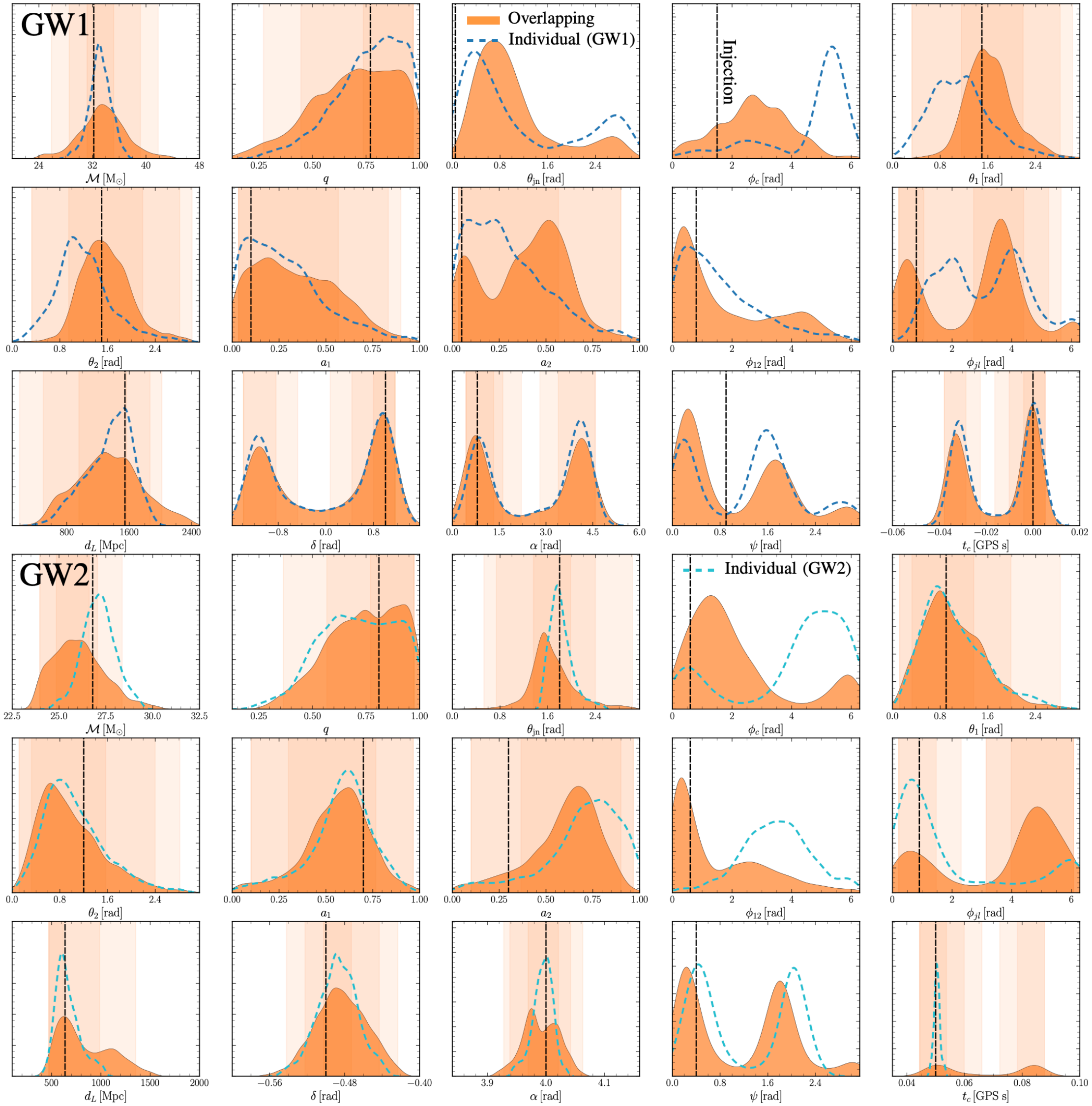}
    \caption{Full set of $\mathrm{1D}$ marginals for our case study with $\Delta t_c = 0.05\,\mathrm{s}$ (\textbf{C1}, as shown in Fig.~\ref{fig:C1_observation}). Inference results from \texttt{peregrine} joint analysis are shown in orange whereas the corresponding single signal analyses are shown in blue and cyan respectively (in order of $t_c$), as shown in the figure legend. The black dashed lines indicate the injected parameter values. For parameters that are well measured, the $1\,\sigma$, $2\,\sigma$ and $3\,\sigma$ contours are indicated in subsequently lighter shading.}
    \label{fig:C1_1D_full}
\end{figure*}

\begin{figure*}
    \centering
    \includegraphics[width=\textwidth,trim={0.1cm 0.1cm 0.1cm 0.1cm},clip]{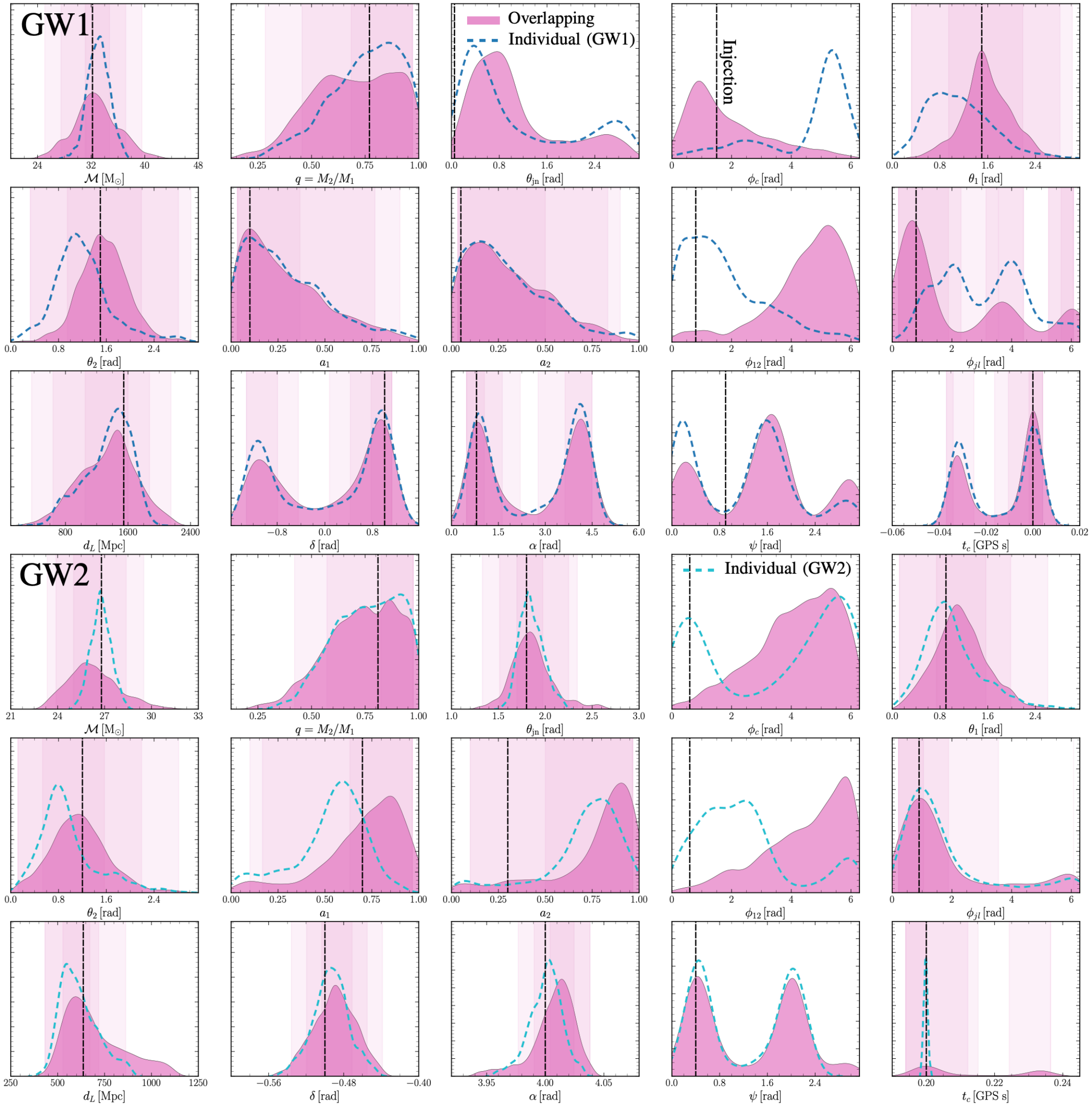}
    \caption{Full set of $\mathrm{1D}$ marginals for our case study with $\Delta t_c = 0.2\,\mathrm{s}$ (\textbf{C2}). Inference results from \texttt{peregrine} joint analysis are shown in pink whereas the corresponding single signal analyses are shown in blue and cyan respectively (in order of $t_c$), as shown in the figure legend. The black dashed lines indicate the injected parameter values. For parameters that are well measured, the $1\,\sigma$, $2\,\sigma$ and $3\,\sigma$ contours are indicated in subsequently lighter shading.}
    \label{fig:C2_1D_full}
\end{figure*}

\begin{figure*}
    \centering
    \includegraphics[width=\textwidth,trim={0.1cm 0.1cm 0.1cm 0.1cm},clip]{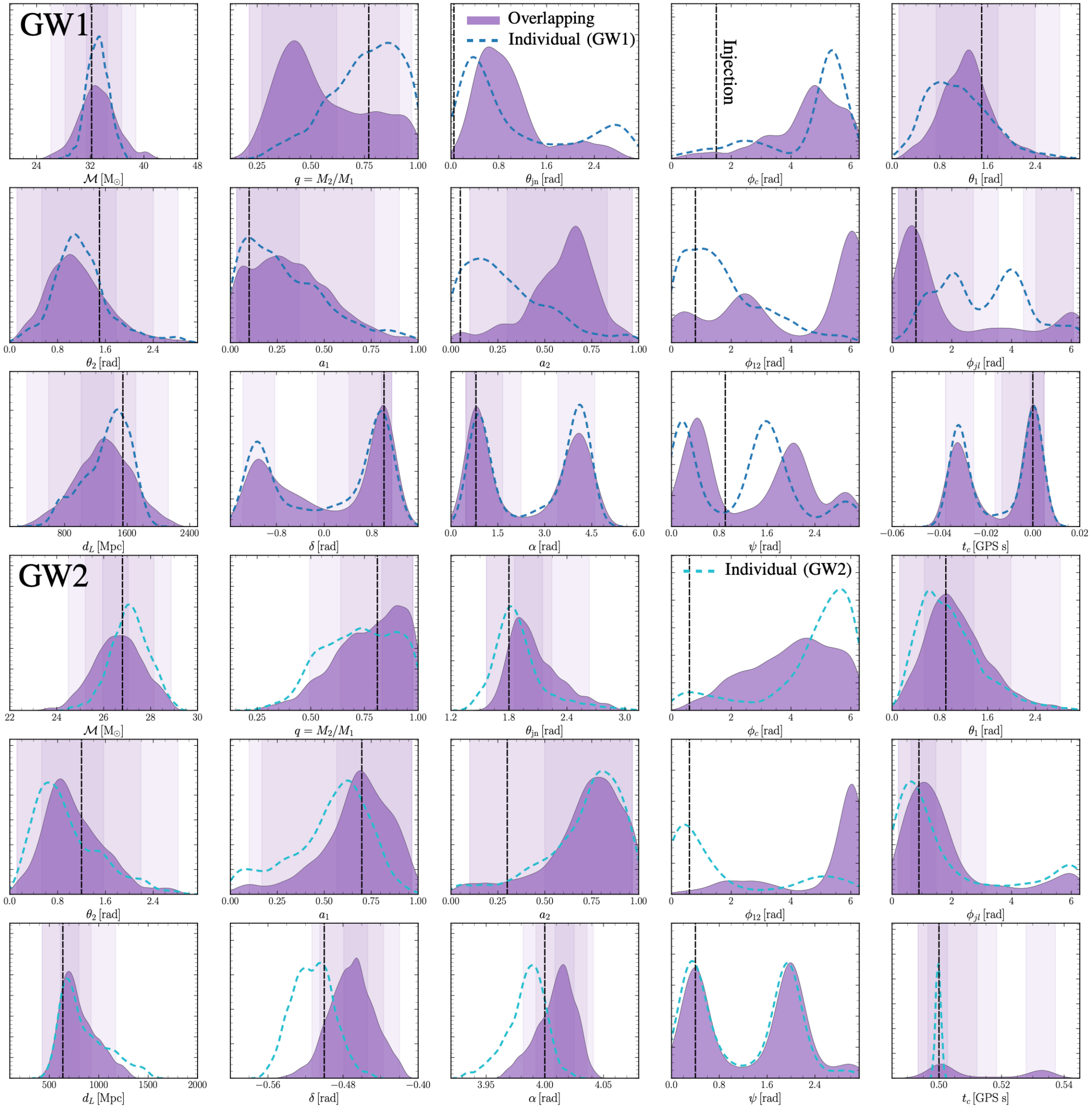}
    \caption{Full set of $\mathrm{1D}$ marginals for our case study with $\Delta t_c = 0.5\,\mathrm{s}$ (\textbf{C3}). Inference results from \texttt{peregrine} joint analysis are shown in purple whereas the corresponding single signal analyses are shown in blue and cyan respectively (in order of $t_c$), as shown in the figure legend. The black dashed lines indicate the injected parameter values. For parameters that are well measured, the $1\,\sigma$, $2\,\sigma$ and $3\,\sigma$ contours are indicated in subsequently lighter shading.}
    \label{fig:C3_1D_full}
\end{figure*}

\begin{figure*}
    \centering
    \includegraphics[width=\textwidth,trim={0.1cm 0.1cm 0.1cm 0.1cm},clip]{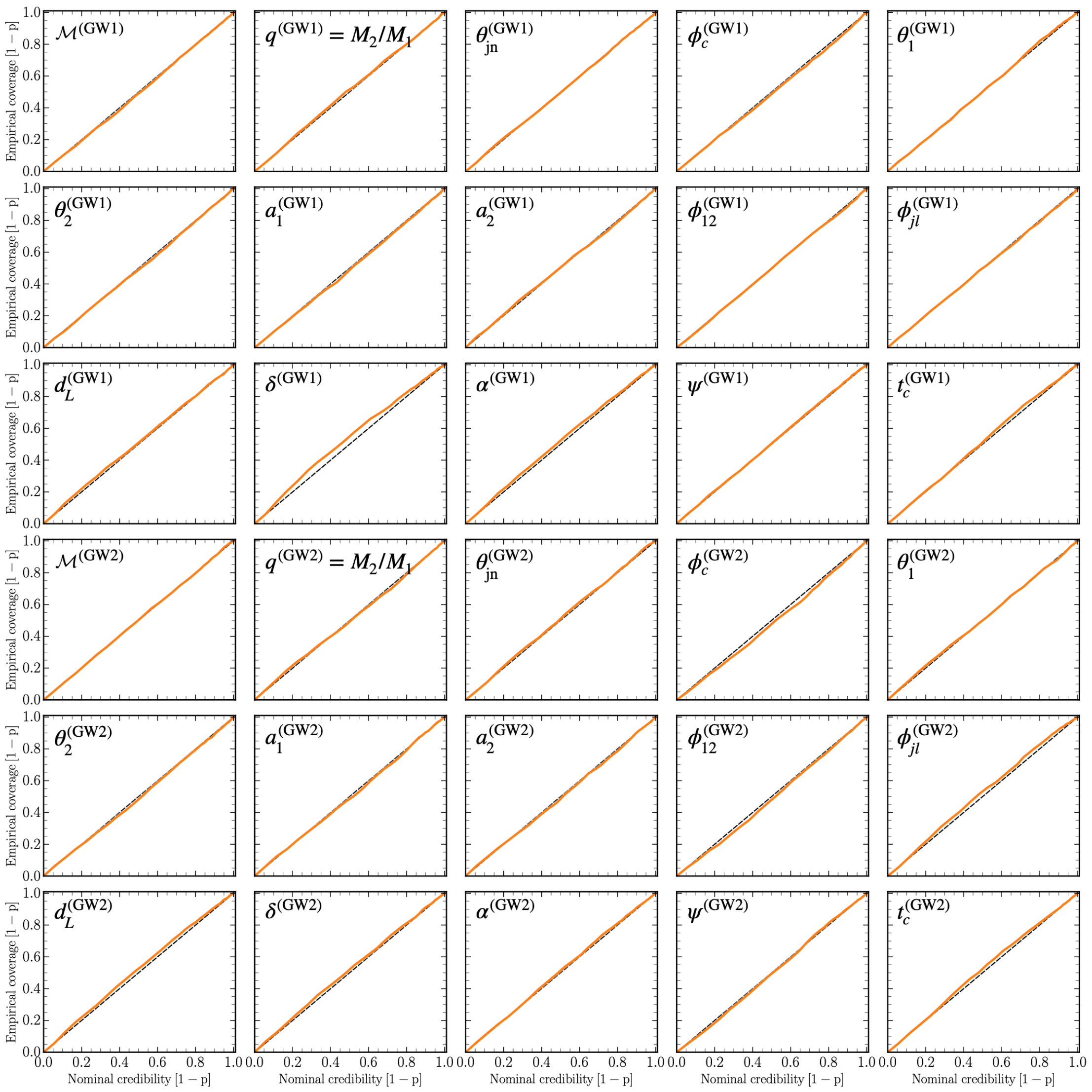}
    \caption{Coverage results for \textbf{C1} for all $30$ parameters characterising an overlapping signal comprising two overlapping binary black hole gravitational wave signals. This compares the expected coverage of the true value as a percentage on the x-axis against the actual coverage of our ratio estimator on the y-axis. The orange lines indicate the average coverage and their strong alignment along the diagonal (black-dashed) shows that our posteriors are well-calibrated.}
    \label{fig:C1_pp_full}
\end{figure*}

\begin{figure*}
    \centering
    \includegraphics[width=\textwidth,trim={0.1cm 0.1cm 0.1cm 0.1cm},clip]{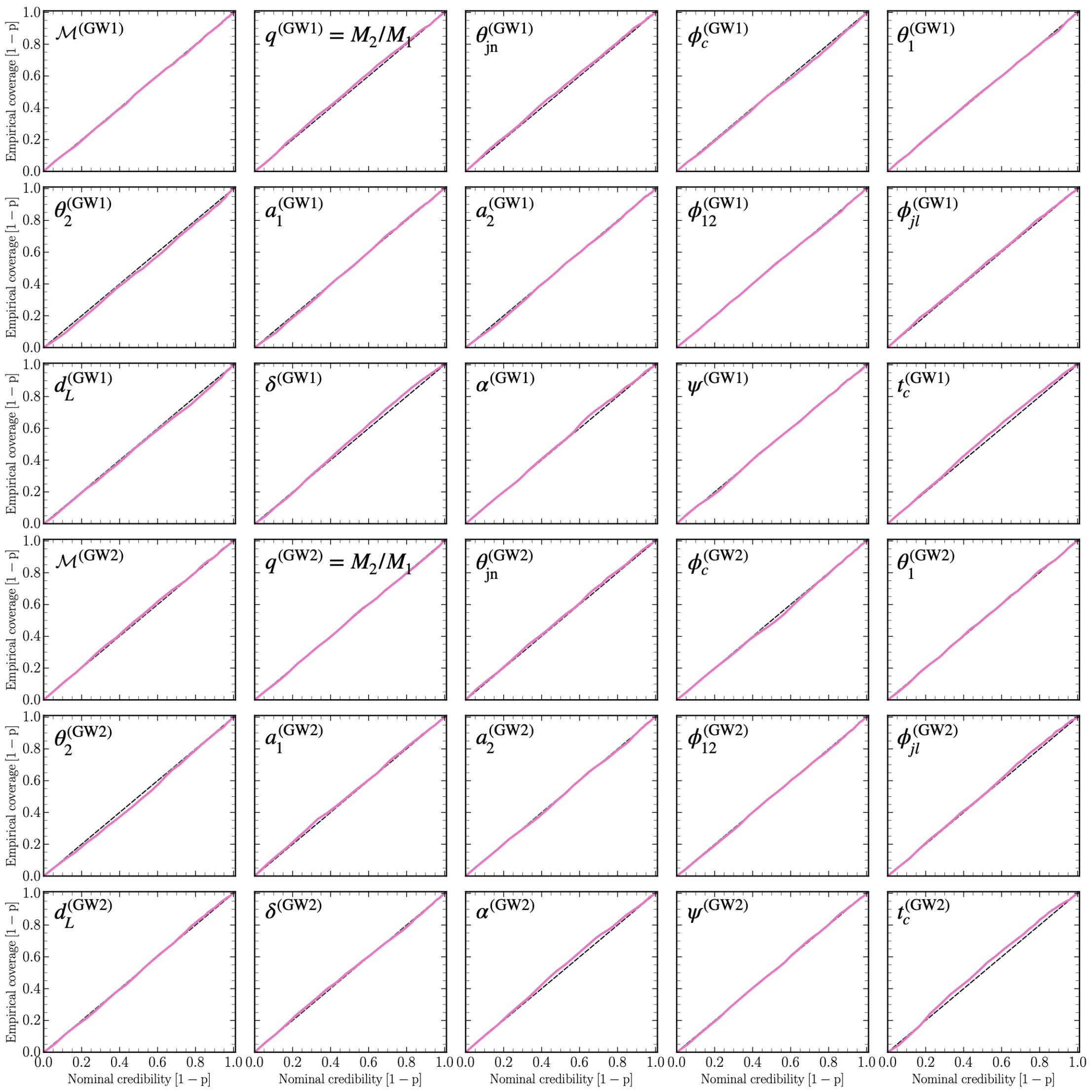}
    \caption{Coverage results for \textbf{C2} for all $30$ parameters characterising an overlapping signal comprising two overlapping binary black hole gravitational wave signals. This compares the expected coverage of the true value as a percentage on the x-axis against the actual coverage of our ratio estimator on the y-axis. The pink lines indicate the average coverage and their strong alignment along the diagonal (black-dashed) shows that our posteriors are well-calibrated.}
    \label{fig:C2_pp_full}
\end{figure*}

\begin{figure*}
    \centering
    \includegraphics[width=\textwidth,trim={0.1cm 0.1cm 0.1cm 0.1cm},clip]{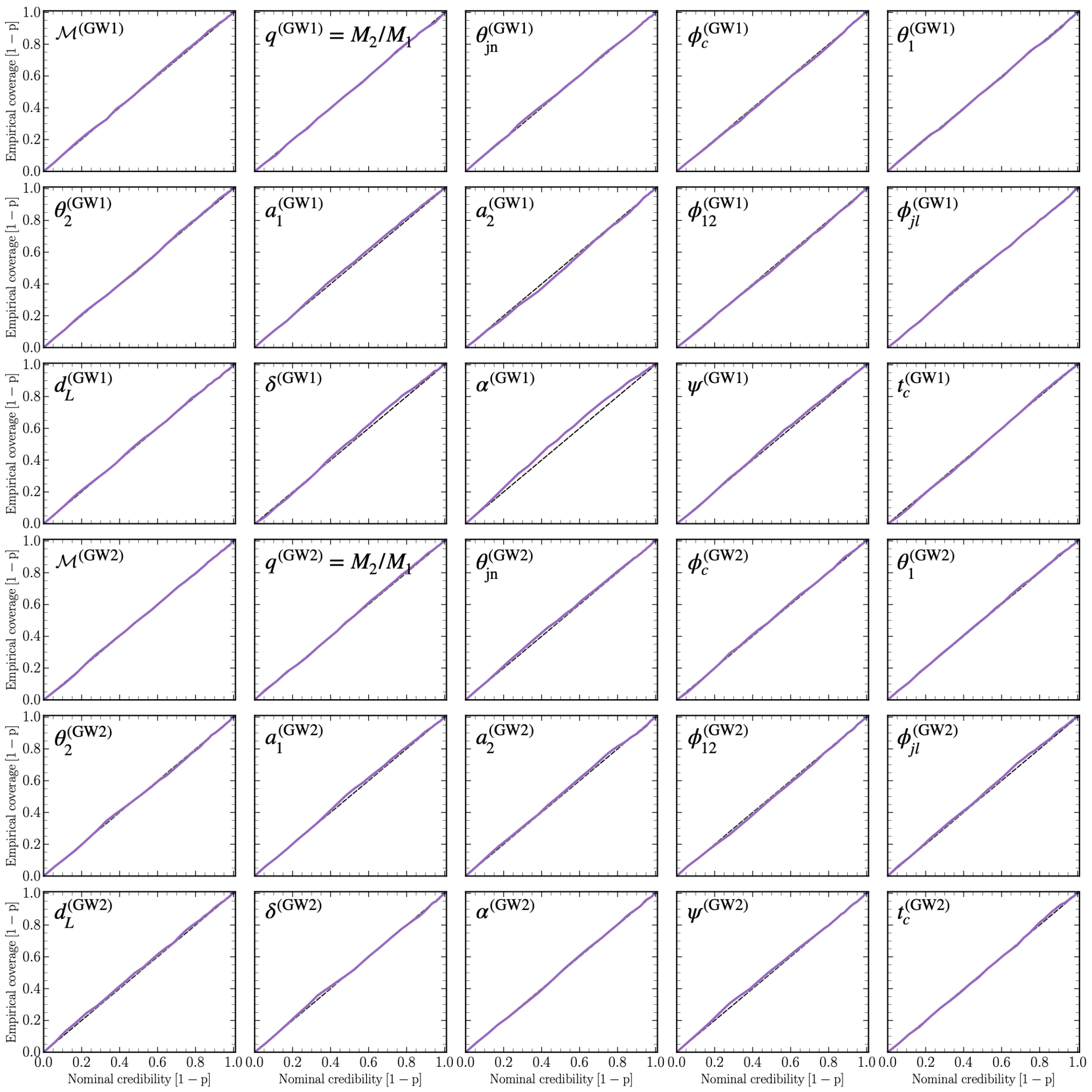}
    \caption{Coverage results for \textbf{C3} for all $30$ parameters characterising an overlapping signal comprising two overlapping binary black hole gravitational wave signals. This compares the expected coverage of the true value as a percentage on the x-axis against the actual coverage of our ratio estimator on the y-axis. The purple lines indicate the average coverage and their strong alignment along the diagonal (black-dashed) shows that our posteriors are well-calibrated.}
    \label{fig:C3_pp_full}
\end{figure*}
\end{document}